\documentclass[article,prx,twocolumn,aps]{revtex4-2}
\usepackage{color}
\usepackage{mathtools}
\usepackage{latexsym}
\usepackage{graphicx}
\usepackage{float}
\usepackage{dcolumn}
\usepackage{bm}
\usepackage{amssymb}
\usepackage{amsmath}
\usepackage{mathtools}
\usepackage[space]{grffile}
\usepackage{xcolor}

\usepackage[colorinlistoftodos]{todonotes}

\newcommand{\be}{\begin{equation}}
\newcommand{\ee}{\end{equation}}
\newcommand{\bea}{\begin{eqnarray}}
\newcommand{\eea}{\end{eqnarray}}

\begin{document}


\title{Critical properties of the prethermal Floquet Time Crystal}

\author{Muath Natsheh$^{1}$}
\author{Andrea Gambassi$^{2,3}$}
\author{Aditi Mitra$^{1}$}
\affiliation{$^{1}$Center for Quantum Phenomena, Department of Physics, New York
University, 726 Broadway, New York, NY, 10003, USA\\
  $^{2}$SISSA - International School for Advanced Studies, via Bonomea 265, 34136
Trieste, Italy\\
$^{3}$INFN - Sezione di Trieste, via Bonomea 265, 34136, Trieste, Italy}

\date{\today}

\begin{abstract}
The critical properties characterizing the formation of the Floquet time crystal
in the prethermal phase
are investigated analytically in the periodically driven $O(N)$ model.
In particular, we focus on the critical line separating the trivial phase with period synchronized dynamics and absence of long-range spatial order from the non-trivial phase where long-range spatial order is accompanied by period-doubling dynamics.
In the vicinity of the critical line,  with a combination of dimensional expansion and exact solution for $N\to\infty$, we determine the exponent $\nu$ that characterizes the divergence of the spatial correlation length of the equal-time correlation functions, the exponent $\beta$ characterizing the growth of the amplitude of the order-parameter, as well as the initial-slip exponent $\theta$ of the aging dynamics when a quench is performed from deep in the trivial phase to the critical line.
The exponents $\nu, \beta, \theta$ are found to be identical to those in the absence of the drive.
In addition, the functional form of the aging is found to depend on whether the system is probed at times that are small or large compared to the drive period. The spatial structure of the two-point correlation functions, obtained as a linear response to a perturbing potential in the vicinity of the critical line, is found to show algebraic decays that are longer ranged than in the absence of a drive, and besides being period-doubled, are also found to oscillate in space at the wave-vector $\omega/(2 v)$, $v$ being the velocity of the quasiparticles, and $\omega$ being the drive frequency.
\end{abstract}
\maketitle

\section{Introduction}

Floquet time crystals (FTCs) are systems that show spontaneous breaking of discrete time-translation symmetry  (TTS),
accompanied by the breaking of another internal symmetry of the system such as an
Ising symmetry \cite{Sacha2018,Else2019,Khemani2019}.
Simply broken Ising symmetry would imply the well known transition from a paramagnetic to a ferromagnetic Ising phase \cite{Goldenfeld92}.
However, the appearance of broken TTS adds a new flavor to this problem.
Thus
natural questions that arise are: Is there any universality associated with the transition from a trivial phase where there is no long-range Ising order, and the dynamics is synchronized with the drive, to a FTC phase? And if yes, what are the critical exponents of the transition? Are they related to those
encountered at the Ising transition in the absence of drive, or are they different?
This question is particularly relevant due to the numerous microscopically different experimental platforms that have realized this phenomenon \cite{Zhang2017,Choi2017,Barrett2018,Volovik2018,Sohan18,Smits18}.

Some of these questions have been addressed in spatial dimension $d=1$ (1$d$) \cite{Yao2017,Berdanier18a,Yates18,Yao2017}. In particular, in the presence of disorder, the system can become many-body-localized in 1$d$ making the FTC, also known
as the discrete time crystal, stable even at long times \cite{Khemani2019}. For a many-body-localized Floquet Ising chain, it was shown that the
critical behavior near the FTC critical point belongs to the infinite randomness universality class \cite{Yao2017,Berdanier18a, Berdanier18b}.  Thus, although driven, the underlying critical behavior could be related to an undriven problem, namely that of a static Ising model with disorder \cite{Fisher92,Fisher94,Fisher95,Damle02,Motrunich00,Vosk15}.

In the absence of disorder, the critical point associated with the FTC of
free fermion Floquet chains, in which exactly
one band is occupied and the others empty, was shown to exhibit a central charge, which was extracted from the scaling of the entanglement entropy \cite{Yates18,Berdanier18a}.
As in equilibrium, this central charge counts the number of Fermi points appearing
at the critical point. With driving, however, new Fermi-points can appear at the Floquet zone boundaries, and the central charge for Floquet chains was found to keep track of these additional Fermi points \cite{Yates18}.

The nature of the transition from a trivial phase to a FTC phase is also a very interesting question for periodically driven Hamiltonian systems coupled to a reservoir, where dissipation and noise make the dynamics effectively classical. For the particular case of the 1$d$ driven-dissipative model studied in Ref.~\onlinecite{Yao2020}, the transition was mapped to the locked-to-sliding transition of a d.c.-driven charge density wave.

The question of universality of the FTC critical point has not been addressed in
spatial dimension $d>1$, for driven but otherwise isolated quantum systems. For $d>1$, even with disorder, one is more likely
to encounter
a prethermal FTC as the fate of many-body-localization above $d=1$ is unknown. The subject of the current paper is the universality
associated with prethermal FTC phases in $d >2$.
Since any Floquet system has a time-independent Floquet Hamiltonian associated with it, it is expected \cite{Else2017} that its critical properties may have
some relation to those of an undriven model. Generically
the undriven model is not the same as the original one with
the driving switched off. This is because the Floquet driving affects non-trivially the parameters of the system, for example, by making the coupling constants more long ranged. In addition, Floquet driving can also change the underlying symmetries.

Non-perturbative approaches such as the large-$N$ limit
have proven to be very useful in understanding
diverse systems ranging from impurity models \cite{Read83,Ratiani09, Mitra11}, Kondo lattice models \cite{Read84}, and strange metals \cite{SY93, Kitaev15}.
In this paper, we address the question of universality in the context of the transition to a FTC by studying the periodically driven $O(N)$ model. In the absence of the drive,
and depending on the value of $N$, this model is a textbook example for studying the transition to various broken symmetry phases both in equilibrium~\cite{Eyal1996, Moshe2003}, and out of equilibrium, due to,  e.g., a quantum quench \cite{Sotiriadis2009,Sotiriadis2010,Sciolla2011,Sciolla2013,Chandran2013,Schmalian2014,Schmalian2015,Tavora2015,Maraga2015, Smacchia2015,Maraga2016,Chiocchetta2016,Lemonik2016,Chiocchetta2017}.
In addition, the model can be exactly solved in the limit $N\to \infty$, providing access to equilibrium and non-equilibrium collective behaviors and critical properties beyond perturbation theory \cite{Moshe2003, Maraga2015, Lemonik2016}.

This paper is organized as follows. In Section \ref{sec:model} we present the model, outline the setup of the problem, and summarize the Gaussian results. In  Section \ref{sec:oneloop}
we present the perturbative one-loop calculation and discuss its effect on the phase diagram and on the correlation length. In Sections \ref{sec:nu} and \ref{sec:beta} we determine the exact expressions of the exponents $\nu$ and $\beta$ in the limit $N\to \infty$ and to $O(\epsilon)$ where $\epsilon= 4-d$, $d$ being the spatial dimension. In Section \ref{sec:aging} we present results for aging following a quench, also in the limit $N\rightarrow \infty$ and to $O(\epsilon)$. In Section \ref{sec:spatial} we investigate the spatial structure of the correlation functions that are obtained as a linear response to a perturbing potential. We present our conclusions in Section \ref{sec:conclu}.
Intermediate details of the calculations are reported in several Appendices.

\begin{figure}[ht]
\includegraphics[width = 0.50\textwidth]{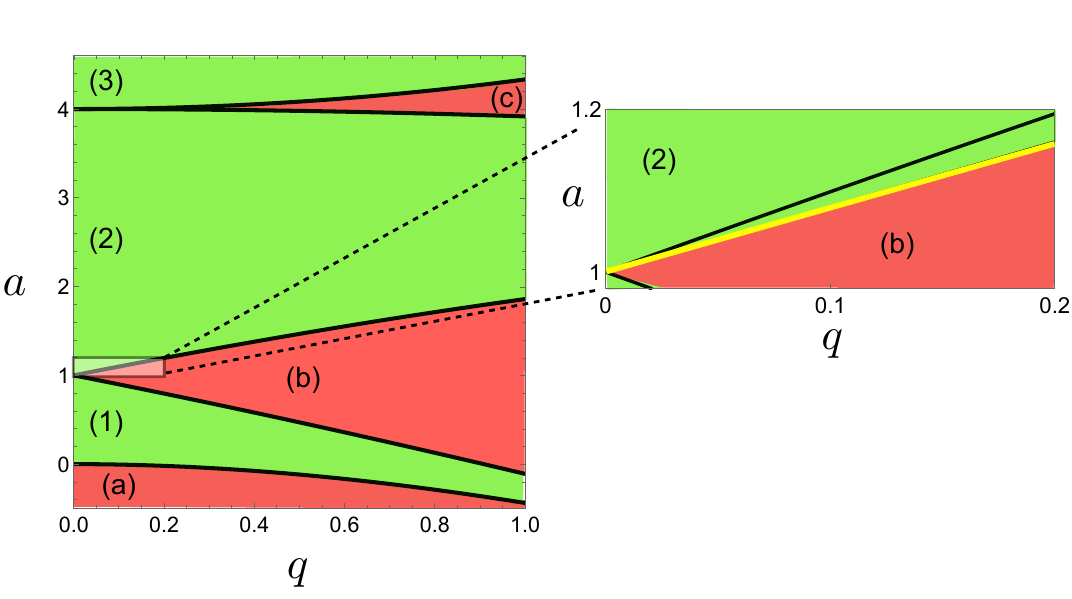}
\caption{\label{fig1}
Stability phase diagram of the periodically driven $O(N)$ model in the $a$-$q$ plane where $a= 4(r+k^2)/\omega^2$ and $q=2r_1/\omega^2$. Compared to the Gaussian approximation reported in the diagram on the left \cite{Natsheh2020}, interactions shift the critical line from $a_1\approx 1+q$  (black line between regions (2) and (b) on the left diagram) to $a_1 \approx 1+q -3\widetilde{B}_4/\omega^2$ (yellow line on the right diagram), for  $q\ll 1$, where $\widetilde{B}_{d}$ is defined in Eq.~\eqref{B0til} and given in Eq.~\eqref{B4def} for $d=4$.}
\end{figure}

\section{Model and set-up} \label{sec:model}

The periodically driven $O(N)$ model we consider is
\begin{align}
H=\sum_{i=1}^{N} \int d^d x \frac{1}{2}\bigl[&(r-r_1\cos{(\omega t)})\phi_i^2(\textbf{x})
+(\vec{\nabla}\phi_i)^2\nonumber\\ &+\Pi_i^2(\textbf{x})\bigr] + V,\label{H}
\end{align}
where \(\phi_i\) and \(\Pi_i\) are \(N\)-component bosonic fields that obey the canonical commutation relation
\begin{equation}
[\phi_j(\textbf{x}),\Pi_l(\textbf{y})]=i\delta_{jl}\delta^{(d)}(\textbf{x}-\textbf{y}).
\end{equation}
\(V\) is the interaction term
\begin{align}
V = \frac{u}{4 N} \int d^d x \biggl(\sum_{i=1}^N \phi_i^2\biggr)^2.\label{V}
\end{align}
In Eq.~\eqref{H}, $\omega = 2\pi/T$ is the frequency of the drive with period $T$, $r_1$ is the drive amplitude, and  we also define the dimensionless drive amplitude
\begin{align}\label{qdef}
q = \frac{2 r_1}{\omega^2}.
\end{align}
In addition, $r$ is a detuning parameter which, for negative values, causes an instability
in the free, undriven model (i.e., with $V=r_1=0$).
Note that the Hamiltonian in Eq.~\eqref{H} has a ${\mathbb Z}_2$ symmetry under $\phi \rightarrow -\phi$ and it also has a discrete TTS, i.e., $H(t+T) = H(t)$. While a conventional Ising ferromagnet corresponds to
broken ${\mathbb Z}_2$ symmetry, we are interested in the FTC phase where both ${\mathbb Z}_2$ and the discrete TTS are spontaneously broken.

For later convenience, we represent the fields
$O_i \in \{ \phi_i,\Pi_i\}$ in momentum space according to
\begin{align}
  O_i(\textbf{x})=\int^{\Lambda}\! \frac{d^d k}{(2\pi)^d} e^{i\textbf{k}.\textbf{x}}O_{i,\textbf{k}},
\label{phik}
\end{align}
where $\Lambda$ is the large-momentum cut-off of the model, which will be  implemented as specified further below.
In the absence of interactions, i.e., within the Gaussian approximation $V=0$, each mode $k$ is independent of the others and its dynamics is given by the Mathieu equations \cite{Nist,McLachlin1947,Richards1983}. The solutions of
these equations is characterized by the dimensionless drive amplitude $q$ introduced in Eq.~\eqref{qdef} and by
another dimensionless quantity
\begin{align}
    a = 4\frac{r + k^2}{\omega^2}. \label{adef}
\end{align}
The solutions of the Mathieu equation result in the stability diagram reported in Fig.~\ref{fig1}, consisting of allowed bands (green regions) and band-gaps (red regions). In the limit of weak drive $q \rightarrow 0$,
the band edges are determined by the condition of parametric resonance for the longest wavelength mode \cite{Natsheh2020}, i.e., integer multiples of the drive frequency $\omega$ should equal the energy for creating a pair of excitations at $k=0$, i.e., $n \omega = 2 \sqrt{r}$,
where $n$  is an integer and $\sqrt{r+k^2}$ is the excitation energy of a mode of wavelength $k$ in the undriven system.
For generic values of $q$, the band edges are given by the Mathieu characteristic values \cite{Nist,McLachlin1947,Richards1983} $a_n(q)$ and $b_n(q)$, where the former corresponds to the upper boundaries and the latter to the lower boundaries of the unstable red regions of Fig.~\ref{fig1}. Note that $a_{n}(q \rightarrow 0) = n^2$ with $a_1(q) = 4 r_c/\omega^2$, $r_c$ being the critical value of $r$ that tunes the system to the upper edge of the unstable region (b) in Fig.~\ref{fig1}.

In the presence of interactions, the phase-diagram of the model, before the onset of heating (which is controlled by
$N$ and $\Lambda$), was discussed in Refs.~\cite{Chandran2016,Natsheh2020}.
In the $q$-$a$ plane, it comprises of a trivial phase with no
spatial long-range order and no period doubling, separated by a series of critical lines to either period synchronized ferromagnetic phases (regions (a) and (c) in Fig.~\ref{fig1}, for $V=0$), or to period-doubled FTC phases (region (b), for $V=0$). Here we discuss the critical properties in the vicinity of one of the FTC critical lines (line separating regions (2) and (b)) which corresponds to the Mathieu characteristic value $a_1(q)$ when $V=0$.

With drive, the emerging critical properties of Eq.~\eqref{H} were discussed in \cite{Natsheh2020} for $V=0$, i.e., within the Gaussian approximation.
It was shown that algebraic behaviors may appear in two-point correlation functions as a prelude to universality. Here we explore the effect of interactions on them. In particular, we combine a perturbative approach with an exact solution in the limit $N\to\infty$, finding that
these algebraic behaviors are robust, that the associated exponents are indeed universal and that they turn out to be identical to those which emerge in the undriven model when tuned to its critical point.

For a weak drive $q\ll 1$, and within the Gaussian approximation, the value $r_c(q)$ of the parameter $r$ corresponding to the critical line is given by  $r_c=(\omega/2)^2 a_1(q)$ with $a_1(q) \approx  1+ q$  \cite{Natsheh2020},
and it is therefore convenient to introduce the parameter
\begin{align}
  y =r-r_c \approx r- \left(\frac{\omega}{2}\right)^2\left(1+q\right), \label{ydef}
\end{align}
which controls the detuning away from criticality.

In our analysis, we consider a quantum quench
\cite{calabrese2007,Mitra2018} in which the initial state is the
thermal equilibrium state of the undriven \(O(N)\) model (i.e., \(r_1=0\)) with a positive initial value $r_0>0$ of the detuning $r$ from the corresponding critical line.
This state is subsequently evolved under the  periodically driven \(O(N)\) model in Eq.~\eqref{H}.
We choose the initial value \(r_0\gg r>0\) so that the initial state is deep in the paramagnetic phase with short-range spatial correlations.
We also choose $r_0\gg u\langle \phi^2\rangle$, where $\langle\phi^2\rangle$ is the average in the initial state. This condition ensures that interactions do not modify this initial state.

Central objects in our study are the two-point correlation functions of the $\phi$ fields, defined by the retarded and Keldysh Green's functions \cite{Kamenevbook} $G_R$ and $G_K$, respectively,
\begin{align}
\delta_{lj}\delta_{\textbf{k},-\textbf{q}}iG_{K}(k,t,t') &=
\langle \left\{\phi_{l,\textbf{k}}(t),\phi_{j,\textbf{q}}(t') \right\}\rangle,\\
\delta_{lj}\delta_{\textbf{k},-\textbf{q}}iG_{R}(k,t,t') &= \vartheta(t-t')
\langle \left[\phi_{l,\textbf{k}}(t),\phi_{j,\textbf{q}}(t') \right]\rangle,
\end{align}
where $\vartheta(t)$ is the step function which is non-vanishing only for $t>0$ and equals one.
Above, since we will be approaching the critical line from the symmetric phase where all the $N$ components of the order-parameter behave equivalently, for notational brevity,
we do not indicate the field components which $G_{K,R}$ refer to.
The Green's functions obtained within the Gaussian approximation will be denoted below by $G_{0K}$ and $G_{0R}$.

It is instructive to write the Keldysh path integral in terms of the $N$-component classical ($\phi_{i,c}$) and quantum fields ($\phi_{i,q}$). In terms of the vectors $\boldsymbol{\phi}_{b}=\left( \phi_{1,b},\phi_{2,b}, ..., \phi_{N,b} \right)$ with $b\in \{c,q\}$, the Keldysh action $\mathcal{S}$ turns out to be
\begin{align}\label{SK}
  \mathcal{S}&=S_{\rm initial} + \int_{\mathbf{x} ,t>0}\biggl\{\dot{\boldsymbol{\phi}}_{q}\cdot \dot{\boldsymbol{\phi}}_{c}
    -\vec{\nabla} \boldsymbol{\phi}_{q}\cdot \vec{\nabla} \boldsymbol{\phi}_{c}
    \nonumber\\
    &\quad \quad - \left[r-r_1\cos(\omega t)\right] \boldsymbol{\phi}_{q}\cdot \boldsymbol{\phi}_{c}\biggr\}+ S_{\rm int},
\end{align}
where $\int_{\mathbf{x} , t}\equiv \int d^d x dt$ and all information about the pre-quench Hamiltonian enters in
$S_{\rm initial}$. Since the quench takes place at $t=0$, all time integrals run over $t>0$.
The contribution $S_{\rm int}$ of the interactions is
\begin{equation}
  S_{\rm int} = -\frac{u}{2N} \int_{\mathbf{x},t>0}\!\!\! \boldsymbol{\phi}_c\cdot\boldsymbol{\phi}_q \left( |\boldsymbol{\phi}_c|^2
  + |\boldsymbol{\phi}_q|^2 \right).
\end{equation}

\subsection{Correlations in the Gaussian approximation}

Under the initial conditions mentioned above, and for the Gaussian theory in which the fields at different wavevectors decouple,
the Green's functions are given by \cite{Natsheh2020}
\begin{align}
  iG_{0K}(k,t,t') =& q^2\frac{\sqrt{r_0}}{2\overline{\omega_k}^2} \cos\left(\frac{\omega}{2}t\right) \cos\left(\frac{\omega}{2}t'\right)
  \nonumber\\ &\times [\cos(\overline{\omega_k}(t-t'))-\cos(\overline{\omega_k}(t+t'))],\label{gkpp}\\
  G_{0R}(k,t,t') =&-\vartheta(t-t')q \cos\left(\frac{\omega}{2}t\right) \cos\left(\frac{\omega}{2}t'\right) \nonumber\\
&\times \frac{\sin(\overline{\omega_k} (t-t'))}{\overline{\omega_k}}, \label{grpp}
\end{align}
where
\begin{align}
  \overline{\omega_k} = \sqrt{\bar{k}^2 +  q y/2}, \quad  {\rm with}\quad \bar{k}= \sqrt{q/2}\, k.\label{ydet}
\end{align}
As expected, on approaching the FTC phase, the unequal time correlators $G_{0R,0K}$ above show period doubling due to the $\cos(\omega t/2)$ prefactors.
On the critical line $y=0$, algebraic prefactors $\bar{k}^{-1}$ and  $\bar{k}^{-2}$ appear in $G_{0R}$ and $G_{0K}$, respectively, which imply an emerging light-cone with quasiparticle velocity $v=\sqrt{q/2}$.

The spatial Fourier transform of $G_{R,K}$ is
\begin{align}
&    G_{R,K}(x,t,t') = \frac{1}{(2\pi)^{d/2} x^{d/2-1}}\nonumber\\
&\quad\quad \quad \times \int_0^{\Lambda} \!\!dk\, k^{d/2}  J_{d/2-1}(k x)\ G_{R,K}(k,t,t'),\label{eq:ft}
\end{align}
where $J_{\alpha}$ is the  Bessel function of the first kind which emerges because of the rotational invariance of the integrand around the direction of $\vec{x}$. Eq.~\eqref{eq:ft} reveals that in spatial dimension $d=4$, $G_{0K}(x,t,t)$ has the following behavior
\cite{Natsheh2020}
\begin{subequations}
\label{gkx}
\begin{align}
  iG_{0K}(x \gg 2 v t) &\approx 0,\\
  iG_{0K}(x=2 v t) &\propto \cos^2\left(\frac{\omega}{2}t\right)\frac{1}{x^{3/2}},\\
  iG_{0K}(x \ll 2 v t) & \propto
\cos^2\left(\frac{\omega}{2}t\right)\frac{1}{x^2}.
\end{align}
\end{subequations}
Accordingly, $G_{0K}$ decays algebraically as $1/x^{3/2}$ upon increasing $x$ on the light-cone $x=2 v t$, as $1/x^2$ inside the light-cone $2 v t \gg x$, while it rapidly vanishes outside the light-cone $x \gg 2 vt$.
Note also that $G_{0K}(k,t,t')$ at equal times $t=t'$ does not show period doubling, but it is synchronized with the drive.

The response function $G_{0R}(x,t,t')$, on the other hand, has an almost delta-function weight on the light-cone \cite{Natsheh2020}, as it vanishes both inside and outside the light-cone while it features an algebraic decay upon increasing $x$ along the light cone $x=v|t-t'|$:
\begin{subequations}\label{grx}
\begin{align}
G_{0R}(x \gg v |t-t'|) &\approx 0,\\
G_{0R}(x=v |t-t'|) & \propto
\cos\left(\frac{\omega}{2}t\right) \cos\left(\frac{\omega}{2}t'\right)\frac{1}{x^{3/2}},\\
  G_{0R}(x \ll v |t-t'|) & \approx 0.
\end{align}
\end{subequations}

\section{Perturbative correction at one loop}
\label{sec:oneloop}

In this section we determine the perturbative one-loop correction to $G_{R,K}$ and discuss its effect on the phase diagram of the model and on its exponent $\nu$.

In particular, the predictions presented here are derived at the lowest order in the coupling constant $u$ and they can be easily extended to account for a generic value of the number $N$ of the  components of the field.
However, we will be eventually interested in the limit $N\to\infty$ and therefore we focus directly on this case.
As the fixed-point value $u^*$ of the coupling constant $u$ is expected to be of order $\epsilon = 4-d$ for spatial dimensionality $d\lesssim 4$, the final expression of the quantities analyzed here will eventually take the form of a dimensional expansion in $\epsilon$, where terms of order $u^2$ and $u \epsilon$ are neglected. In order to simplify the notation we will explicitly indicate the order of approximation only in the final expressions.

The one-loop corrections $\delta G_{R,K}$ to $G_{R,K}$ can be written as \cite{Tavora2015,Chiocchetta2016}
\begin{subequations}\label{dG}
\begin{align}\label{dGr}
\delta G_{R}(k,t,t')=\int_{0}^\infty d\tau\, G_{0R}(k,t,\tau)\;iT(\tau)\;G_{0R}(k,\tau,t'),
\end{align}
\begin{align}\label{dGk}
 \delta iG_{K}(k,t,t')&=\int_{0}^\infty d\tau\, G_{0R}(k,t,\tau)\; iT(\tau) \; iG_{0K}(k,\tau,t')\nonumber\\
 &\quad\quad\quad\quad+(t\leftrightarrow t'),
\end{align}
\end{subequations}
where $i T(t)$ is the tadpole integral given by
\begin{align}
iT(t)&= u \langle \phi^2\rangle =\frac{u}{2}iG_{0K}(x=0,t,t)\nonumber\\
&=\frac{u}{2}\int\frac{d^dk}{(2\pi)^d} \,iG_{0K}(k,t,t)f_c\left(k/\Lambda\right),
\label{Tad-1}
\end{align}
for $N \rightarrow \infty$.
In the expression above $f_c(x)$ is a cut-off function with $f_c(0)=1$. This function is also  assumed to be smooth and to vanish for $x\gg 1$ in order to reduce the effects of microscopic oscillations in the dynamical quantities and expose possible underlying universal behaviors \cite{Tavora2015,Maraga2015,Chiocchetta2016}.
In Appendix \ref{app-tad} we show that the tadpole in Eq.~\eqref{Tad-1} takes the form
\begin{align}
iT(t)&= \widetilde{B}_d \cos^2(\omega t/2) + iT'(t),\label{Tad}
\end{align}
where $\widetilde{B}_d$ is the constant determined by the leading behavior $iT(t) \propto \cos^2(\omega t/2)$ for $t\to\infty$ while $iT'$ is the time-dependent transient part of the tadpole, which vanishes as $t$ increases.
In $d$ spatial dimension, $\widetilde{B}_d$ is given in
Eq.~\eqref{Bddef} in Appendix \ref{app-tad}, which we report here for
convenience
\begin{align}
\widetilde{B}_d(y) =  a_d
\frac{q u}{2}\sqrt{r_0}\int_0^{\infty}dk\, k^{d-1} \frac{f_c\left(k/\Lambda\right)}{k^2 +y},\label{B0til}
\end{align}
where $a_d = 2/[(4\pi)^{d/2}\Gamma(d/2)]$.
For convenience we choose $f_c(x)=e^{-x}$ but a different choice does not affect the large-distance behavior as long as $f_c(0)=1$ and $f_c(x)$ vanishes sufficiently fast as $x$ increases.
As the tadpole above is $\propto u$ --- and since the fixed-point value  of
$u$, denoted by $u^*$, is proportional to $\epsilon = 4-d$, --- the expression multiplying $u$ can be evaluated directly at $d=4$. With this observation, the transient part of the tadpole evaluated at $d=4$ turns out to be
\begin{align}
  i T'(t) \approx \frac{8 \theta}{3 q t^2}\cos^2(\omega t/2)\quad  {\rm with}\quad  \theta= \frac{3 q u\sqrt{r_0}}{(16\pi)^2}, \label{Tadt}
\end{align}
at times $t\gg \bar{\Lambda}^{-1}$ which are longer than the microscopic scale set by $\bar{\Lambda} = \sqrt{q/2}\Lambda$.

Before considering the effects of the transient contribution to the tadpole on the resulting $G_{R,K}$, we discuss below the effect of the long-time part $\propto {\widetilde B}_d$ in Eq.~\eqref{Tad}.
In particular, in Sec.~\ref{subsec:CL} we discuss the perturbative correction it provides to
the critical line while in Sec.~\ref{subsec:PTnu} we focus on the correction to the correlation length $\xi$ which characterizes the spatial decay of the equal-time correlation functions at long times.

\subsection{Perturbative correction to the critical line}
\label{subsec:CL}

The Keldysh action of the model is given by Eq.~\eqref{SK} and, in the absence of interactions, i.e., with $S_{\rm int} \mapsto 0$,  the part of it which is proportional to the fields $\boldsymbol{\phi}_{q}\cdot \boldsymbol{\phi}_{c}$ is $[r-r_1\cos(\omega t)] \boldsymbol{\phi}_{q}\cdot \boldsymbol{\phi}_{c}$. In the presence of $S_{\rm int}$, this term is corrected as follows at one-loop
\begin{align}
 & \left[r-r_1\cos(\omega t)\right]   \boldsymbol{\phi}_{q}\cdot \boldsymbol{\phi}_{c} \mapsto \nonumber\\
& \quad \left[r-r_1\cos(\omega t) + u \langle \phi^2\rangle\right]\boldsymbol{\phi}_{q}\cdot \boldsymbol{\phi}_{c}.
\end{align}
Using Eqs.~\eqref{Tad-1} and \eqref{Tad} this expression implies the following correction at long times
\begin{align}
 & \left[r-r_1\cos(\omega t)\right]   \boldsymbol{\phi}_{q}\cdot \boldsymbol{\phi}_{c} \mapsto \nonumber\\
& \quad
\left[r-r_1\cos(\omega t) + \widetilde{B}_4 \cos^2(\omega t/2)\right]\boldsymbol{\phi}_{q}\cdot \boldsymbol{\phi}_{c}.
\end{align}
Writing $\cos^2(\omega t/2) = [1+\cos(\omega t)]/2$, it is clear that the effect of the long-time part of the tadpole is to shift
\begin{align}
r\mapsto r + \frac{\widetilde{B}_4}{2} \quad\mbox{and}\quad r_1\mapsto r_1 - \frac{\widetilde{B}_4}{2},\label{rsh}
\end{align}
with the resulting effective action, after these shifts, being still Gaussian at this order in perturbation theory.

Recall that the critical line for the Gaussian model is given by the condition $y=0$ in Eq.~\eqref{ydef}, i.e., the critical value $r_c$ of $r$ for $q \ll 1$ is determined by
\begin{align}
    r_c \approx \frac{\omega^2}{4} + \frac{r_1}{2}. \label{rcdef}
\end{align}
The shifts in Eq.~\eqref{rsh} imply that
the critical value $r_c$ of $r$ is now determined by the condition above imposed on the shifted parameters, i.e.,
\begin{align}
r_c + \frac{\widetilde{B}_4}{2} = \frac{\omega^2}{4} + \frac{r_1}{2}-\frac{\widetilde{B}_4}{4}.
\end{align}
Accordingly, in terms of the parameters $r$ and $r_1$ of the original model and of the associated dimensionless quantities $a$ and $q$ (see Eqs.~\eqref{qdef} and \eqref{adef}),
the critical line is shifted compared to that in the absence of interaction (i.e., with $\widetilde{B}_4 \mapsto 0$) as
\begin{align}
a_1(q) = \frac{4 r_c}{\omega^2} = 1 + q - 3\frac{\widetilde{B}_4}{\omega^2}.
\end{align}
The critical line is therefore shifted downwards, i.e., the stable green disordered region, due to the interaction, widens locally at the expense of the red unstable (ordered) region, as usually occurs also in equilibrium.   This fact is shown schematically in Fig.~\ref{fig1}.

\subsection{Perturbative correction to the exponent $\nu$} \label{subsec:PTnu}

We now discuss the perturbative correction to
the correlation length $\xi$ which characterizes the spatial decay of the equal-time correlation function at long times.
This then allows us to determine perturbatively the associated critical exponent $\nu$ to $O(u)$, which will eventually yield an estimate at the lowest order in the dimensional expansion $\epsilon = 4-d$.
In Section \ref{sec:nu} we will then determine the exact dependence of $\nu$ on the dimensionality $d$ for $N\to\infty$ beyond the perturbative result presented here.

The shifts in Eq.~\eqref{rsh} caused by the long-time limit of the
tadpole imply a shift
\begin{equation}
y \mapsto y + 3 \widetilde{B}_4(y)/4
\label{eq:ysh}
\end{equation}
in the detuning from the critical line introduced in Eq.~\eqref{ydef}.
Based on the dependence of $G_{0R}$ on $k$ in Eq.~\eqref{grpp} (which, is not altered by the contribution of the tadpole beyond the shifts in the parameters in Eq.~\eqref{rsh}), one can easily identify the spatial correlation length $\xi$ of the field as being determined by
$y =\xi^{-2}$.
Accordingly,  the correlation length accounting for the one loop correction is (see Eq.~\eqref{eq:ysh})
\begin{align}
    \xi^{-2} = y + \frac{3}{4} \widetilde{B}_4(y).\label{yg}
\end{align}
Let us denote by $y_c$ the critical value of $y$, which is shifted by the interaction and is determined by the condition that, correspondingly, $\xi$ diverges, i.e., from Eq.~\eqref{yg},
\begin{align}
0 = y_c + \frac{3}{4} \widetilde{B}_4(y_c). \label{ycd}
\end{align}
Subtracting Eq.~\eqref{ycd}
from Eq.~\eqref{yg} one finds
\begin{align}
\xi^{-2} &= y-y_c + \frac{3}{4}\left[\widetilde{B}_4(y) - \widetilde{B}_4(y_c)\right]\nonumber\\
&= y - y_c  + 3\frac{u}{8} a_4 q\sqrt{r_0}\nonumber\\
&\times \int_0^{\infty} dk\, k^3
\biggl[\frac{1}{k^2 + y} -\frac{1}{k^2 + y_c}\biggr]f_c\left(k/\Lambda\right),
\end{align}
where, in the last equality, we used Eq.~\eqref{B0til}.
Denoting by $\delta y = y-y_c$ the distance from the actual critical line, the previous equation can be written as
\begin{align}
\xi^{-2} &= \delta y \biggl[1 - 3\frac{u}{8} a_4 q\sqrt{r_0} \nonumber\\
&\quad\quad \times
\int_0^{\infty}dk  k^3  \frac{f_c\left(k/\Lambda\right)}{\left(k^2+\delta y+ y_c\right)\left(k^2+y_c\right)}\biggr].
\label{eq:xi-int}
\end{align}
Note that, from Eq.~\eqref{ycd},  we expect $y_c$ to vanish in perturbation theory as the interaction strength $u$ vanishes
and therefore we may set $y_c=0$ in the denominator of
the integrand in Eq.~\eqref{eq:xi-int} as the prefactor of the integral is already of order $u$ and the calculation is done at the lowest order in perturbation theory.
By choosing the exponential cut-off function introduced after Eq.~\eqref{B0til} (see also Eq.~\eqref{fcdef})
and after
performing the momentum integral, one finds
\begin{align}
    \xi^{-2} &= \delta y \biggl[1 - 3\frac{u}{16} a_4 q\sqrt{r_0}
    \ln\biggl|\frac{\Lambda^2}{\delta y}\biggr| + O(u^2)
\biggr],\label{xip}
\end{align}
for $\delta y \ll \Lambda^2$.
At this order in the perturbative expansion, this logarithmic correction  can be exponentiated in order to obtain
\begin{align}
\xi^{-2} &= \delta y \biggl[1 + q u A \ln\biggl|\frac{\delta y}{\Lambda^2}\biggr|+O(u^2)\biggr] \nonumber \\
&\approx \Lambda^2 \biggl|\frac{\delta y}{\Lambda^2}\biggr|^{1+ A q u} + O(u^2), \label{nup}
\end{align}
where we introduced
\begin{align}
A \equiv \frac{3}{16} a_4 \sqrt{r_0} =\frac{3}{128 \pi^2} \sqrt{r_0}.
\end{align}
This analysis reveals that the exponent $\nu$, defined by the algebraic singularity of $\xi \approx |\delta y|^{-\nu}$, is modified by the interaction compared to its Gaussian value $\nu =1/2$. The latter is recovered here by setting $u=0$.
Although in Eq.~\eqref{nup} the coupling constant $u$ is determined by the microscopic value entering the model, in order to determine the leading scaling behavior close to the critical line, $u$ is effectively replaced by its renormalization-group fixed-point value $u^*$; we later argue that in the limit $N\to\infty$ this value can be determined and is given by, c.f.,  Eq.~\eqref{ufp}. Accordingly, from Eq.~\eqref{nup}, we conclude that the resulting exponent $\nu$ takes the value
\begin{align}
    \nu = \frac{1+ A qu^*}{2} = \frac{1}{2}\left[1 +\frac{\epsilon}{2} +O(\epsilon^2)\right],
    \label{nufp}
\end{align}
where $\epsilon = 4 -d$.

So far we have considered the effects of introducing the interactions within a perturbative expansion at the lowest order in the coupling constant $u$, eventually leading to expressions which are perturbative in the actual expansion parameter $\epsilon$. While the expansion in $u$ can be easily generalized to the case of finite $N$, and may also be used also to analyze the effects of the transient part of the tadpole, the fixed-point value $u^*$ of $u$ is easily determined only for $N\to \infty$, as shown below. This is the reason why we primarily focus on the limit $N\rightarrow \infty$.
In the next section, we determine the exact exponents $\nu$ and $\beta$ for $N\to\infty$ as functions of the spatial dimensionality $d$.  In particular, the expansion of the expression of $\nu$ for $d = 4 -\epsilon$ at the first order in $\epsilon$ renders, as expected, Eq.~\eqref{nufp}.

\section{Exact solution for $\nu$}
\label{sec:nu}

In the $N \rightarrow \infty$ limit, the evolution equation of $G_K$ derived from the action in Eq.~\eqref{SK} becomes exact and can be solved  self-consistently. Noting that $\langle \phi_q^2\rangle =0$ \cite{Kamenevbook} we obtain
\begin{equation}
\begin{split}
\big[ \partial_t^2 + k^2+ r -r_1 \cos(\omega t) &+ \frac{u}{2}iG_K(x=0,t,t)\big]\\
&\times iG_K(k,t,t')= 0.
\end{split}
\label{self1}
\end{equation}
Motivated by Eqs.~\eqref{Tad-1} and \eqref{Tad}, which show that
$iG_{0K}(x=0,t,t) \propto \cos^2(\omega t/2)$, we make the ansatz that in the equation above one has
\begin{align}
& r -r_1 \cos(\omega t) + \frac{u}{2}iG_K(x=0,t,t) \nonumber\\
&= r+\delta r -(r_1+\delta r_1)\cos(\omega t),\label{self2}
\end{align}
where $iG_K$ is the Keldysh function of the Gaussian model ($u=0$)
but calculated with the renormalized parameters
$r+\delta r$ and $r_1+\delta r_1$ instead of $r$ and $r_1$, respectively.
These new parameters also imply that the effectively Gaussian detuning from the critical line, determined from Eq.~\eqref{ydef}, takes the form
\begin{align}
    y = r+\delta r - \frac{r_1+\delta r_1}{2} - \frac{\omega^2}{4}.\label{ydef1a}
\end{align}
Substituting Eq.~\eqref{gkpp} in Eq.~\eqref{self2}, and using the long-time limit of
the tadpole in Eq.~\eqref{Tad}, we find that the self-consistency condition requires
\begin{equation}
\delta r = \widetilde{B}_d/2 \quad\mbox{and} \quad \delta r_1 = - \widetilde{B}_d/2,
\end{equation}
where $\widetilde{B}_d$ is given in Eq.~\eqref{B0til}.
Accordingly, the detuning in Eq.~\eqref{ydef1a} becomes
\begin{align}
y = r - \frac{r_1}{2} -\frac{\omega^2}{4} +  \frac{3}{4}\widetilde{B}_d(y).
\label{y-ex}
\end{align}
This equation provides the implicit relationship between the Gaussian detuning $y$ and the control parameters $r$ and $r_1$.
We also note here that within this effective Gaussian model with renormalized parameters, the spatial correlation length $\xi$ is still related to $y$ as $y=\xi^{-2}$.
Since the critical line corresponds to $y=0$, the critical values $r_{c}$ and $r_{1c}$ of the parameters $r$ and $r_1$, respectively, satisfy Eq.~\eqref{y-ex} with $y=0$ (equivalently $\xi = +\infty$), i.e., $0 = r_c - r_{1c}/2 -\omega^2/4  + (3/4)\widetilde{B}_d(0)$.
Subtracting this condition from Eq.~\eqref{y-ex}, and using Eq.~\eqref{B0til}, we find the self-consistency condition in terms of the correlation length $\xi$, i.e.,
\begin{align}
  \xi^{-2} = r-r_c -\frac{r_1-r_{1c}}{2} + g \Lambda^{d-2}\biggl[\omega_d\left(\Lambda^{-1}/\xi\right) - \omega_d(0)\biggr],
  \label{eq:xi-self-c}
\end{align}
where we introduced
\begin{align}
  g = \frac{3}{8} q u\sqrt{r_0}, \label{gdef}
\end{align}
and the function $\omega_d(z)$ which captures the scaling behavior of $\widetilde{B}_d(\xi^{-2})$ as
\begin{equation}
\int \frac{d^d k}{(2\pi)^d}\frac{f_c\left(k/\Lambda\right)}{k^2 + \xi^{-2}} = \Lambda^{d-2} \omega_d\left(\Lambda^{-1}/\xi\right).
\label{def-omega-d}
\end{equation}
We emphasize here that Eq.~\eqref{eq:xi-self-c} turns out to be independent of the driving frequency $\omega$ which influences, instead, the location of the critical line. In fact, Eq.~\eqref{eq:xi-self-c} has the same structure as the equation which controls the correlation length $\xi$ in the undriven model, both after the quench \cite{Chiocchetta2016} and in equilibrium  \cite{Moshe2003}.
In the latter context it was shown in Ref.~\cite{Moshe2003} that
\begin{align}
\omega_d(z) = \omega_d(0) -K_d z^{d-2} + c_d z^2 + \mbox{h.o.},
\label{exp-omega_d}
\end{align}
where $c_d$ is a constant that depends on the cut-off function $f_c$ in Eq.~\eqref{def-omega-d},
while $K_d$ is a universal constant given by
\begin{align}
  &K_d =a_d \int_0^{\infty} dz \frac{z^{d-1}}{z^2(1+z^2)}=-a_d\frac{\pi/2}{\sin(\pi d/2)},
\end{align}
with $K_d>0$ for $2 < d < 4$.
When $d<4$, the contribution from $c_d z^2$ is subleading compared to $K_d z^{d-2}$ for $z\to 0$. Accordingly, for $\xi \gg \Lambda^{-1}$, the term in brackets on r.h.s.~of Eq.~\eqref{eq:xi-self-c} is more relevant than that on the l.h.s.~and the equation implies that
\begin{equation}
\xi \propto \left(r-r_c -\frac{r_1-r_{1c}}{2}\right)^{-\nu},
\label{eq:def-nu}
\end{equation}
with
\begin{align}
    \nu = \frac{1}{d-2}.
    \label{eq:nu-d}
\end{align}
This exponent $\nu$ controls the divergence of the spatial correlation length of the fluctuations within the system as $r-r_1/2 \to r_c - r_{1c}/2$, i.e., as the critical line in Fig.~\ref{fig1}(b) is approached.
For $d> 4$, instead, the term $\propto z^2$ in Eq.~\eqref{exp-omega_d} is the dominant one for $z\to 0$ and therefore in Eq.~\eqref{eq:xi-self-c}  the contributions $\propto \xi^{-2}$ on the l.h.s.~and r.h.s. are the relevant ones. Accordingly, one finds the same expression as in Eq.~\eqref{eq:def-nu}  with
\begin{align}
\nu = \frac{1}{2}.
\end{align}

The predictions presented above for the leading scaling behavior and for the exponent $\nu$ are actually independent of the coupling constant $g$ of the model which can take any (positive) value. However, among these values, there is a specific one $g^*$ for which the leading corrections to the scaling behavior discussed above do vanish. In the renormalization-group framework, $g^*$ correspond to the fixed-point value of $g$, see, e.g., Ref.~\cite{Moshe2003}.
In particular, for $2<d<4$, the corrections to the scaling in Eq.~\eqref{eq:def-nu} come from the term $\propto z^2$ in Eq.~\eqref{exp-omega_d} and from the analogous one on the l.h.s.~of Eq.~\eqref{eq:xi-self-c}. Taking them into account, the consistency condition in Eq.~\eqref{eq:xi-self-c} can then be written as
\begin{align}
\xi^{-2}\biggl[1 - g c_d \Lambda^{d-4}\biggr]= r - r_c -\frac{r_1-r_{1c}}{2}- g K_d \xi^{2-d}.
\end{align}
The expression of $\xi$ which can be derived from this equation matches the scaling behavior in Eq.~\eqref{eq:def-nu}, i.e., with no leading corrections to scaling, only if the l.h.s.~of the equation vanishes, i.e., for $g=g^*$, where
\begin{equation}
g^* = \Lambda^{4-d}/c_d.
\end{equation}
While $c_d$ is a non-universal constant which depends on the cutoff function $f_c$ (see Eqs.~\eqref{exp-omega_d} and \eqref{def-omega-d}), it turns out \cite{Moshe2003} that its limit for $d\to 4$, i.e., $\epsilon \to 0$, is universal and given by $c_{4-\epsilon} = 1/(8 \pi^2 \epsilon) + O(\epsilon^0)$,
which renders
\begin{align}
g^* = 8 \pi^2 \epsilon + O(\epsilon^2),
\label{gst-eps}
\end{align}
in agreement with what is expected from renormalization-group arguments.
This fixed-point value $g^*$ of the coupling constant $g$ in the dimensional expansion can now be used in the perturbative expansions discussed in Sec.~\ref{sec:oneloop} to set the fixed-point value of the coupling constant $u$ which, using Eq.~\eqref{gdef}, is given by
\begin{align}
u^* = \frac{(8\pi)^2 }{3 q \sqrt{r_0}}\epsilon + O(\epsilon^2).
\label{ufp}
\end{align}
In particular, as anticipated in Sec.~\ref{subsec:PTnu}, the perturbative prediction for $\nu$ can be recovered from the expression in Eq.~\eqref{nup} if we substitute $u$ by its fixed point value $u^*$, and assume $\epsilon \ll 1$. This leads to Eq.~\eqref{nufp}, which indeed coincides with the dimensional expansion of Eq.~\eqref{eq:nu-d} for $d=4-\epsilon$ at the first order in $\epsilon$. This is also similar to what happens in equilibrium.

In Sec.~\ref{sec:aging} the perturbative fixed-point value $u^*$ of $u$ determined here will be used to provide a prediction for the scaling exponents which emerge in perturbation theory when studying the short-time behavior of $G_{R,K}$.

\section{Exact Solution for $\beta$} \label{sec:beta}

In order to extract the exponent $\beta$ that governs the behavior of the order parameter close to criticality, we assume that the symmetry of the model is (spontaneously) broken along one spatial direction, i.e., that  $\langle \phi_i(t)\rangle =\sqrt{N}M(t)\delta_{i,1}$ for $i=1,\ldots, N$.
Then for $N \rightarrow \infty$, the equations of motion for $M$ and $G_K$ are as follows \cite{Chiocchetta2016, Chandran2016}:
\begin{subequations}\label{HFM}
\begin{align}
&\biggl[ \partial_t^2 + r -r_1 \cos(\omega t) + u M^2(t) \nonumber\\
&\qquad+ \frac{u}{2}iG_K(x=0,t,t)\biggr]M(t) = 0,\\
&\biggl[ \partial_t^2 + k^2+ r -r_1 \cos(\omega t) + u M^2(t) \nonumber\\
&\qquad + \frac{u}{2} iG_K(x=0,t,t)\biggr]iG_K(k,t,t') = 0.
\end{align}
\end{subequations}
Assuming that the magnetization $M$ shows period doubling, we make the ansatz $M(t) = M_0 \cos(\omega t/2)$  and find that the self-consistent equations written above imply that (see Appendix \ref{app-betaex} for details)
\begin{equation}\label{Msol}
|M_0| \propto \sqrt{r_c-r+ r_{1c}/2 - r_{1}/2},
\end{equation}
and therefore
\begin{align}\label{betaeq}
\beta =1/2,
\end{align}
as for the model in equilibrium \cite{Moshe2003}.

\section{Aging following a quench}
\label{sec:aging}

In this section we discuss how the correlation functions behave in the transient regime following a quantum quench. For this, we consider the transient part of the tadpole given in Eq.~\eqref{Tadt} and we substitute it into Eqs.~\eqref{dGr} and \eqref{dGk}, which are derived at the lowest order in perturbation theory.
We recall that in Secs.~\ref{subsec:PTnu}, \ref{sec:nu}, and \ref{sec:beta} we focused on the long-time behavior by neglecting this transient.
A key quantity that turns out to characterize this early-time regime is the constant $\theta$, defined perturbatively on the basis of the transient behavior $iT'(t)$ of the tadpole in Eq.~\eqref{Tadt}. As discussed below, this constant is eventually the exponent which appears in the scaling of $G_{R,K}$.
By using the perturbative fixed-point value $u^*$ of the coupling $u$ in Eq.~\eqref{ufp}, $\theta$ in Eq.~\eqref{Tadt} takes the value
\begin{align}
\theta^*= \frac{\epsilon}{4} + O(\epsilon^2).
\label{thfp}
\end{align}
This exponent $\theta^*$ is identical to the one obtained for the undriven model \cite{Tavora2015, Maraga2015,Chiocchetta2016}, in which the higher-order corrections $O(\epsilon^2)$ turn out to vanish as $N\rightarrow \infty$ for $2<d<4$, providing an exact exponent \cite{Maraga2015}.
In this section we discuss how this $\theta$ appears in the driven problem.

In the transient regime, for $k=0$ and $t\gg t'$, but with both times longer than the drive period $T= 2\pi/\omega$, we find that $G_R= G_{0R} +\delta G_R$ becomes (see Appendix \ref{app-grtr} for details)
\begin{align}
 &G_{R}(k=0,t \gg t', \omega t \gg  \omega t'\gg 1 )  \nonumber\\
  &\quad=-q t \cos(\omega t/2)\cos(\omega t'/2)\left[1-\theta\ln\bigl(t/t'\bigr)\right]\nonumber \\
  &\quad\approx -q t \cos(\omega t/2)\cos(\omega t'/2)\bigl(t'/t\bigr)^{\theta}.\label{agr1}
\end{align}
Using Eq.~\eqref{thfp}, this implies that at the fixed point, the universal
exponent $\theta^*$ governs the aging dynamics, which manifests itself via the dependence of $G_R$ on $(t'/t)^{\theta^*}$.
On the other hand for $t \gg t'$, but with one time being smaller than the drive period $T$, and the other longer, i.e., for $\omega t \gg 1 \gg \omega t'$,
one finds  (see Appendix \ref{app-grtr} for details)
\begin{align}
& G_{R}(k=0,t \gg t', \omega t\gg 1 \gg \omega t' ) = -q t \cos(\omega t/2)\nonumber\\
&\qquad\quad \times \cos(\omega t'/2)
\biggl[1-\theta\ln{(\omega t)} +  \frac{8}{3} \theta \ln{(\omega t')}\biggr]\nonumber \\
  &\qquad \approx -q t \cos(\omega t/2)\cos(\omega t'/2) (\omega t')^{8 \theta/3}/(\omega t)^{\theta}.\label{agr2}
\end{align}
Equations~\eqref{agr1} and \eqref{agr2} show that, due to the appearance of two exponents, $\theta$ and $ 8\theta /3$, the functional form of the observed aging, and the exponents which control it, are modified depending on whether the system is probed at longer or shorter times compared to the drive period.
The resummation of the leading logarithmic behavior done in the previous equation, leading to an algebraic dependence, is consistent within perturbation theory but it certainly needs to be put on a firmer ground by an analytic solution or a full-fledged renormalization-group analysis of the problem, as was done in the absence of driving \cite{Maraga2015,Chiocchetta2016,Tavora2015}.

In a similar manner, the
Keldysh Green's function in perturbation theory at one-loop, when probed at times longer than the drive period, turns out to be (see Appendix \ref{app-gktr} for details)
\begin{align}
  & i G_K(k=0, \omega t\gg 1,\omega t' \gg 1 )  \nonumber\\
&\quad  =  q^2 \sqrt{r_0} \cos(\omega t/2)\cos(\omega t'/2)
\; t t'\biggl[1 -
    \theta \ln\bigl(\bar{\Lambda}^2 t t'\bigr)\biggr]\nonumber\\
 &\quad \propto q^2 \sqrt{r_0}
\cos(\omega t/2)\cos(\omega t'/2)  \bigl(\bar{\Lambda}^2 t t'\bigr)^{1 - \theta}.
\label{gktr1}
\end{align}
On the other hand, for $\omega t \gg 1\gg \omega t'$, one finds
(see Appendix \ref{app-gktr})
\begin{align}
&  i G_K(k=0, \omega t \gg 1 \gg \omega t') = q^2 \sqrt{r_0} \cos(\omega t/2)\nonumber\\
&\times \cos(\omega t'/2)  t t'\biggl[1 -
    \theta\ln(\bar{\Lambda} t) - \frac{8}{3} \theta \ln(\bar{\Lambda}t')\biggr]\nonumber\\
  &\propto q^2 \sqrt{r_0}  \cos(\omega t/2)\cos(\omega t'/2)
  (\bar{\Lambda}t)^{1- \theta} (\bar{\Lambda}t')^{1-8 \theta/3}.\label{gktr2}
\end{align}
Thus here too --- with the same proviso as the one spelled out after Eq.~\eqref{agr2} on the resummation of the logarithmic dependence ---  the perturbative expressions indicate that the aging behavior changes depending on whether the time is small or large compared to the period of the drive.

Now we discuss the behavior of $iG_K(k,t,t)$ for $kt \gg 1$. Within the Gaussian approximation in
Eq.~\eqref{gkpp} this quantity decays as $1/\bar{k}^2$ upon increasing $\bar{k}$. The one loop corrections slow down this decay to $1/\bar{k}^{2-3\theta}$.
In particular we have (see Appendix \ref{app-gklt} for details)
\begin{align}
iG_K(k, t, t) \approx q^2
\cos^2(\omega t/2)\frac{\sqrt{r_0}}{\bar{k}^{2-2 \theta}}\sin^2(\bar{k} t).\label{gktr3}
\end{align}

The expressions above indicate that, for a quantum quench to the FTC critical line, the correlators at long wavelengths $\bar{k}\ll \omega$ and with one of the times longer than the drive period, i.e., $\omega t \gg 1$, can be cast in the following scaling forms, valid within the prethermal regime and reminiscent of what was found in the undriven model \cite{Tavora2015,Maraga2015,Chiocchetta2016},
\begin{align}
    G_K(k,t,t') &=\cos(\omega t/2)\cos(\omega t'/2) \frac{1}{k^{2-2\theta}}{\cal G}_K(kt,kt',\omega t'),\\
    G_R(k,t,t')&=\cos(\omega t/2)\cos(\omega t'/2)\frac{\bigl(t'/t\bigr)^{\theta}}{k}
    {\cal G}_R(kt,kt',\omega t').
\end{align}
In order for these expressions to render the behaviors discussed above, we assume ${\cal G}_{R,K}(x,y,z) \sim 1$ for $x$, $y$, $z \gg 1$. For  $x$, $y\ll 1$, instead,
${\cal G}_K(x,y,z) \sim (x y)^{1-\theta} f^{-1}(z)$, with $f(z)\sim 1$ for $z\gg 1$ and $f(z)\sim z^{5\theta/3}$ for $z\ll 1$. In the remaining case $y\ll x\ll 1$, one has ${\cal G}_R(x,y,z) \sim x f(z)$.

\begin{figure}[ht]
\includegraphics[width = 0.45\textwidth]{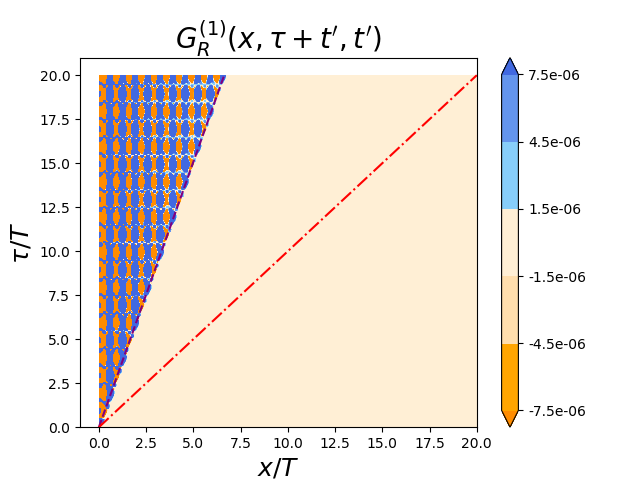}
\caption{\label{figGR}
One-loop correction \(G_R^{(1)}(x,\tau+t',t')\) arising from
the perturbation in Eq.~\eqref{Vpdef} in $d=4$, in the $(\tau,x)$ plane for fixed value of $t' = 0.1$, $r_0=1$, and dimensionless drive amplitude $q=0.22$. We take the strength of the perturbation to be $a_p=1$, while $\tau$ and $x$ are measured in units of the drive period $T$.
The dashed line indicates the light-cone with
quasiparticle velocity $v=\sqrt{q/2}\simeq 0.33$ while the
dot-dashed line corresponds to $v=1$ \cite{Natsheh2020}, which characterizes the light-cone in the absence of the drive.
The spatial oscillations occur with a characteristic wave-vector $\omega/(2 v)$, while the asymptotic decay of  \(G_R^{(1)}(x,\tau+t',t')\) at large distances is described by Eq.~\eqref{GR1x}.}
\end{figure}

\section{Spatial structure due to  resonances}\label{sec:spatial}

We now discuss the effects on the dynamics of adding a perturbing potential of the form
\begin{align}
  V_p(t) = a_p\left[1-2\cos(\omega t)\right], \label{Vpdef}
\end{align}
with $a_p/\omega^2 \ll 1$. This particular choice of $V_p$ does not change the value of  $r-r_1/2$ compared to the unperturbed case $V_p=0$ and thus, from Eq.~\eqref{ydef}, the detuning from the critical line is not affected.
Accordingly, the Floquet quasi-modes for $a_p\neq 0$ will lead to a behavior which is still described by the Gaussian correlators in Eqs.~\eqref{gkpp} and \eqref{grpp}.
However, here we address a different question: assuming that the
perturbation $V_p$ was switched on suddenly at time $t=0$, what is the linear response to this perturbation of a state which is initially at the critical line and therefore described by the correlators in Eqs.~\eqref{gkpp}, \eqref{grpp} with $y=0$?

\begin{figure}[ht]
\includegraphics[width = 0.45\textwidth]{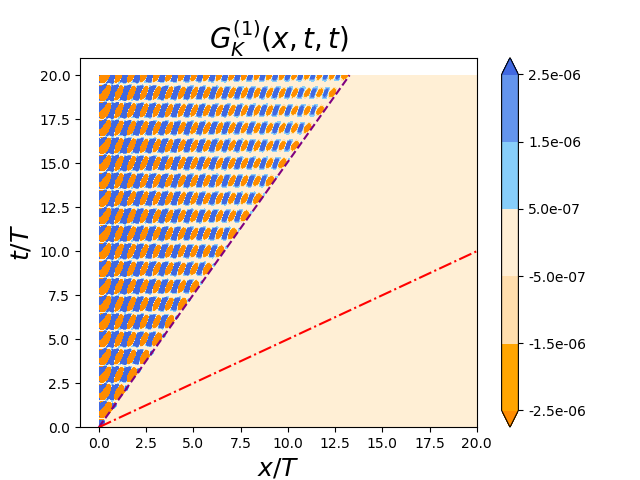}
\caption{\label{figGK}
One-loop correction \(G_K^{(1)}(x,t,t)\) in the $(t,x)$ plane, for the same values of the parameters as in Fig.~\ref{figGR}. Both space $x$ and time $t$ are expressed in units of the period $T$ of the drive.  The light-cone (dashed line) is located at $x=2 v t$, with $v=\sqrt{q/2}\simeq 0.33$. For comparison, we also report the line corresponding to $v=1$ (dot-dashed line), which characterizes the light-cone in the absence of the driving.
The correlator decays as $x^{-3/2}$ upon increasing $x$ both on and inside the light-cone and it oscillates in space with the typical wave-vector $\omega/(2 v)$. This correlator, like all the equal-time correlators does not show period doubling.}
\end{figure}

Treating $V_p$ perturbatively,  the one-loop correction to the  correlators, given by Eqs.~\eqref{dGr} and  \eqref{dGk}, with $iT(\tau)\mapsto V_p(\tau)$, are found to be
\begin{subequations}\label{dGR1}
\begin{align}
G_R^{(1)}(k, t,t') &= -\cos(\omega t/2)\cos(\omega t'/2)\frac{q^2 a_p}{8 \bar{k}^2}\!\!\!\sum_{m=\pm 1,\pm 2}
\!\!\!I^R_{m \omega},\\
iG_K^{(1)}(k,t,t)&=
\cos^2(\omega t/2)\frac{q^3\sqrt{r_0}a_p}{8 \bar{k}^3}\!\!\sum_{m=\pm 1,\pm 2}I^K_{m\omega},
\end{align}
\end{subequations}
where $G^{(1)}_{R,K}$ are the one-loop corrections to the Gaussian parts already reported in Eqs.~\eqref{gkpp}, \eqref{grpp}, and
\begin{subequations}\label{IRK}
\begin{align}
    &I^R_{m \omega}(k,t,t') =\frac{\cos(m\omega (t+t')/2)\sin((\bar{k}-m\omega/2) (t-t'))}{\bar{k}-m\omega/2}\nonumber\\
    &\qquad\quad-\cos(\bar{k}(t-t'))\biggl[\frac{\sin(m\omega t)-\sin(m\omega t')}{m\omega}\biggr],\\
&I^K_{m \omega}(k,t) =\frac{[1+\cos(m \omega t)][1-\cos(2\bar{k}t)]}{2\bar{k}-m\omega}
 \nonumber \\
&\qquad\quad-\sin(2\bar{k}t)\sin(m\omega t)\biggl[\frac{1}{m\omega}+\frac{1}{2\bar{k}-m\omega}\biggr].
\end{align}
\end{subequations}
Although $G^{(1)}_{R,K}$ have the same form as the Gaussian correlators in Eqs.~\eqref{gkpp} and \eqref{grpp}, for long wavelengths $\bar{k}/\omega \ll 1$, the spatial Fourier transform in Eq.~\eqref{eq:ft}, which we perform for $d=4$, is sensitive to the presence of resonances for $k=m\omega/(2v)$, with $m = \pm 1, \pm 2$.
These resonances modify drastically the dependence of $G^{(1)}_{R,K}$ on space,  as shown in Figs.~\ref{figGR} and \ref{figGK}. As expected, a light-cone is visible in $G^{(1)}_R(x,t,t')$ for $x=v|t-t'|$ and in the equal-time correlator $iG^{(1)}_K(x,t,t)$ for $x=2 v t$.
However, the one-loop corrections to $G^{(1)}_{R,K}$ upon increasing the distance $x$ are characterized by a decay in space which is slower compared to the Gaussian correlators.
Moreover, $G^{(1)}_{R,K}$ are found to oscillate in space with the wave-vector $m \omega/2v$ with $m=1,2$.
The spatial oscillations are clearly visible in Figs.~\ref{figGR} and \ref{figGK} for $G_R^{(1)}$ and $G_K^{(1)}$, respectively.

The asymptotic form of $G_{R,K}^{(1)}$ on the light cone and inside it turn out to be (see Appendix \ref{app-ft} for details)
\begin{subequations}\label{GR1x}
\begin{align}
&G_R^{(1)}(x\leq v|t-t'|) \propto \cos(\omega t/2)\cos(\omega t'/2)\nonumber\\
&\quad\times \sum_{m=1,2}\cos\biggl(m\frac{\omega (t+t')}{2}\biggr)\frac{\sin(m \omega x/(2 v) +\delta)}{x^{3/2}},\\
&G^{(1)}_K(x \leq 2 v t, t,t) \propto \cos^2(\omega t/2)
\sum_{m=1,2}\frac{\cos(m\omega x/2v +\delta')}{x^{3/2}},
\end{align}
\end{subequations}
where $\delta$ and $\delta'$ are phase-shifts.
These expressions should be compared with the corresponding ones
in the unperturbed state (i.e., with $a_p=0$), which are given in Eqs.~\eqref{gkx} and \eqref{grx}.
In particular, on the light-cone one finds that the correlators $G^{(1)}_{R,K}$ decay as $x^{-3/2}$ upon increasing $x$, which is also the case for the Gaussian correlators on the light-cone. However, the spatial decay of $G^{(1)}_{R,K}$ is also accompanied by oscillations in space, as is clearly visible in Figs.~\ref{figGR} and \ref{figGK} when looking along the light-cone (dashed lines).

Inside the light-cone, the difference between  the behavior of $G^{(1)}_{R,K}$ and the Gaussian correlators $G_{0R,0K}$, are more pronounced. For example while $G_{0R}$ vanishes inside the light-cone, the one loop correction to it, $G^{(1)}_R$, is non-zero, and decays as $x^{-3/2}$. In a similar manner while $G_{0K}$ decays as $x^{-2}$ inside the light-cone, the one-loop correction to it, $G^{(1)}_K$ decays more slowly, as $x^{-3/2}$. All these decays are also accompanied by spatial oscillations at the wave-vectors $m\omega/(2 v)$, with $m=1,2$.

%
%
%
%
%
%
%

\section{Conclusions} \label{sec:conclu}

In this paper we have studied the physics of a
prethermal  Floquet time crystal (FTC) in the vicinity of the critical line that separates it from a trivial phase. The results are derived for the periodically driven $O(N)$ model with $N\to \infty$. Exact results for the exponents $\nu$ and $\beta$ are derived in the long-time (but still prethermal) regime. Results are also obtained in the transient regime following a quantum quench where the system shows aging dynamics controlled by an exponent $\theta$. The results in this transient regime are obtained  to the leading order in $\epsilon = 4-d$.

The critical exponents $\nu$ (see Eq.~\eqref{eq:nu-d}), $\beta$ (see Eq.~\eqref{betaeq}), and $\theta$  (see Eq.~\eqref{thfp}) are found to be the same as those of the undriven problem, provided that the system is probed at times longer than the drive period $T$ (see Eqs.~\eqref{agr1}, \eqref{gktr1}). At times shorter than $T$, a perturbative treatment reveals that a different scaling behavior possibly emerges
(see Eqs.~\eqref{agr2}, \eqref{gktr2}).

FTCs always arise when the drive is resonant with some microscopic scale of the undriven problem. We showed that this leads to certain peculiarities such as the linear response of the system to a time-dependent perturbing potential can lead to spatial oscillations controlled by the drive frequency and the quasiparticle velocity.

Our study leaves a number of open questions for further investigations. Solving numerically the dynamics of the system near the FTC critical line and  extracting from these solutions the exponents would be a useful exercise. Carrying out a complete renormalization-group analysis which not only renders the exponents predicted here, but generalizes them to finite $N$ is a direction worth pursuing together with an exact analysis of the short-time behavior for $N\to \infty$.

Finally, exploring other universality classes of FTCs, especially those where the exponents $\nu$, $\beta$, and $\theta$ may be different from the undriven situation, is an exciting direction of research. Investigating critical properties even of undriven systems \cite{Wilczek2012_Q,Wilczek2012_C,Li2012}, or non-Hermitian systems \cite{Buca19}, which can nevertheless show time-crystal behavior \cite{Fazio2018,Svistunov2020,Else2017,Else2019,Khemani2019,Kozin2019,Sondhi2020,Kozin2020,Wright19,Sacha20c,Wright20}  despite no-go theorems \cite{Bruno2013_1,Bruno2013_2,Bruno2013_3,Oshikawa2015}, is also an important open question.


{\sl Acknowledgements:}
This work was supported by the US National Science Foundation Grant NSF-DMR 2018358.


\appendix

\section{Tadpole correction}\label{app-tad}
In this section we outline the derivation of Eq.~\eqref{Tad}.
Note that the Green's functions within the Gaussian approximation, in the long wavelength limit $k \ll \sqrt{q} \omega \ll \omega $, and along the critical line $y=0$  are obtained from Eqs.~\eqref{gkpp} and \eqref{grpp} by setting $\overline{\omega_k} = \bar{k}$, see Eq.~\eqref{ydet},
\begin{align}
  iG_{0K}(k,t,t') &= q^2\frac{\sqrt{r_0}}{2\bar{k}^2} \cos\left(\frac{\omega}{2}t\right) \cos\left(\frac{\omega}{2}t'\right)
  \nonumber\\
  & \times [\cos(\bar{k}(t-t'))-\cos(\bar{k}(t+t'))],\label{gkppa}\\
  G_{0R}(k,t,t') &=-\vartheta(t-t')q \cos\left(\frac{\omega}{2}t\right) \cos\left(\frac{\omega}{2}t'\right) \nonumber\\
&\times   \frac{\sin(\bar{k} (t-t'))}{\bar{k}}. \label{grppa}
\end{align}
It is interesting to note that the above Green's functions can be written in terms of
the Green's functions of the undriven problem following a quantum quench \cite{Natsheh2020}, as follows
\begin{subequations}\label{gdu}
\begin{align}
G_{0R}(k,t,t') &= q \cos(\omega t/2)\cos(\omega t'/2)\ G_{0R,u}(\bar{k},t,t'),\label{gdur}\\
iG_{0K}(k,t,t') &= q^2 \cos(\omega t/2)\cos(\omega t'/2)\ G_{0K,u}(\bar{k},t,t'), \label{gduk}
\end{align}
\end{subequations}
where \cite{Tavora2015, Chiocchetta2016} we denote the undriven Green's functions following a quantum quench as $G_{0R,u}, G_{0K,u}$
\begin{subequations}\label{gu}
\begin{align}
G_{0R,u}(k,t,t')&=-\vartheta(t-t')\frac{\sin(k(t-t'))}{k},\\
iG_{0K,u}(k,t,t')&=\frac{\sqrt{r_0}}{2k^2}\bigl[\cos(k(t-t'))-\cos(k(t+t')\bigr].
\end{align}
\end{subequations}

Using Eq.~\eqref{gkppa}, the tadpole in Eq.~\eqref{Tad} can be written as
\begin{align}
  iT(t)&=\frac{u}{2}\int \frac{d^dk}{(2\pi)^d} iG_{0K}(k,t,t) f_c\left(k/\Lambda\right)\nonumber\\
  &= q^2\cos^2(\omega t/2)\frac{u}{2}\sqrt{r_0}\nonumber\\
&\times  \int \frac{d^dk}{(2\pi)^d} \frac{f_c\left(k/{\Lambda}\right)}{2\bar{k}^2}\biggl[1-\cos(2\bar{k}t)\biggr]\label{tadin}\\
  &= \widetilde{B}_d\cos^2{(\omega t/2)} + iT'(t).\label{Tad2}
\end{align}
The first term in the last line corresponds to the long-time behavior of the tadpole
\begin{align}
  \widetilde{B}_d(y) =
 a_d\frac{q u}{2} \sqrt{r_0}\int_0^{\infty} dk \, k^{d-1}  \frac{f_c\left(k/\Lambda\right)}{k^2 +y},\label{Bddef}
\end{align}
with $a_d =2/[(4\pi)^{d/2}\Gamma(d/2)]$. The second term
$iT'(t)$ in Eq.~\eqref{Tad2} denotes the transient behavior
of the tadpole.

We discuss the long-time behavior first.  For spatial dimension $d=4$, and restoring a small detuning $y$, the coefficient $\widetilde{B}_4$ in Eq.~\eqref{Bddef} is given by
\begin{align}
\widetilde{B}_4(y) =
a_4\frac{q u}{2}  \sqrt{r_0}\int_0^{\infty} dk \, k^3  \frac{f_c\left(k/\Lambda\right)}{k^2 +y},\label{B4def}
\end{align}
where $a_4 = 1/(8\pi^2)$.

Now we turn to the leading transient behavior. Removing the contribution from the long-time part, and keeping track of the cut-off $\Lambda$, we have
\begin{align}
  iT'(t;\Lambda)&=\frac{u}{2}\int \frac{d^dk}{(2\pi)^d}iG_{0K}(k,t,t)f_c\left(k/\Lambda\right)\nonumber\\ &\qquad\qquad-\widetilde{B}_d\cos^2(\omega t/2).
\end{align}
We use an exponential cut-off function,
\begin{align}
    f_c(x) = e^{-x}. \label{fcdef}
\end{align}
Note that the long time and long distance behavior is independent of the precise details of the cut-off function.
Using Eq.~\eqref{gduk}, and at spatial dimension $d=4$, we obtain
\begin{align}
 iT'(t;\Lambda) &= -\widetilde{B}_4\cos^2(\omega t/2)+ q^2\cos^2(\omega t/2) \frac{u}{16\pi^2}\nonumber\\
&\times \left(\frac{4}{q^2}\right)\biggl[ \int_0^\infty d\bar{k} \bar{k}^3\exp(-\bar{k}/\bar{\Lambda}) iG_{0K,u}(\bar{k},t,t)\biggr] \nonumber \\
&=\frac{8\theta}{3 q}\cos^2(\omega t/2)\biggl\{ (2\bar{\Lambda})^2 \frac{(2\bar{\Lambda} t)^2-1}{[(2\bar{\Lambda} t)^2+1]^2}\biggr\},\label{Tdu}
\end{align}
where
\begin{align}
\theta= \frac{3 q u\sqrt{r_0}}{256 \pi^2}.
\end{align}
In the long-time limit $\bar{\Lambda} t\gg 1$, one obtains
\begin{equation}\label{tadlt}
iT'(t,\Lambda)\xrightarrow{\bar{\Lambda} t\rightarrow \infty} \frac{8 \theta}{3 q t^2}\cos^2(\omega t/2),
\end{equation}\\
i.e, Eq.~\eqref{Tadt} in the main text.

\section{Exact calculation of $\beta$}
\label{app-betaex}
Let us denote the non-zero magnetization along a certain direction $i$ to be $\langle \phi_i\rangle = \sqrt{N} M$. The equations of motion in the $N\rightarrow\infty$ limit
are in Eq.~\eqref{HFM}. The self-consistent equations are equivalent to the coefficient before the $\boldsymbol{\phi}_{q}\cdot \boldsymbol{\phi}_{c}$ term in the Keldysh action Eq.~\eqref{SK} having the following form
\begin{align}
r - r_1 \cos(\omega t) + u M^2(t)+\frac{u}{2}iG_K(x=0,t,t).
\end{align}
Let us make the ansatz for period doubling
\begin{align}
  M(t) = M_0 \cos(\omega t/2) \Rightarrow M^2(t) = \frac{M_0^2}{2}\left[1+\cos(\omega t)\right].\label{Mod}
\end{align}
In addition, as for the case of $M=0$ discussed in Section~\ref{sec:nu}, we assume that $iG_K$ is the Keldysh function of the Gaussian model ($u=0$)
but calculated with renormalized parameters
$r +\delta r$ and $r_1+\delta r_1$.
Thus, substituting for $iG_K$ from Eq.~\eqref{Tad}, and for $M^2$
from Eq.~\eqref{Mod}, we obtain
the self-consistent equation
\begin{align}
&  r - r_1 \cos(\omega t) + \frac{u M_0^2}{2}\left[1+\cos(\omega t)\right] + \frac{\widetilde{B}_d}{2}
  \left[1+\cos(\omega t)\right]\nonumber\\
&  =  r + \delta r - (r_1 +\delta r_1) \cos(\omega t).
\end{align}
Matching the coefficients on both sides gives
\begin{align}
  r + \frac{1}{2}\widetilde{B}_d(y) + \frac{u M_0^2}{2}&= r + \delta r,\\
  r_1 -\frac{1}{2} \widetilde{B}_d(y)-\frac{u M_0^2}{2} &= r_1 + \delta r_1.
\end{align}
The self-consistent equation for the detuning $y$ follows from Eq.~\eqref{ydef1a}:
\begin{align}
  y = r - \frac{r_1}{2} -\frac{\omega^2}{4} +  \frac{3}{4}\widetilde{B}_d(y) + \frac{3}{4}u M_0^2.\label{M1}
\end{align}
When $y \rightarrow 0$, the above becomes
\begin{align}
  0 =  r - \frac{r_1}{2} -\frac{\omega^2}{4} +  \frac{3}{4}\widetilde{B}_d(0) + \frac{3}{4}u M_0^2.\label{M2}
\end{align}
Recall that the critical couplings $r_c, r_{1c}$ are such that $y=0$ when $M_0=0$, corresponding to the condition
\begin{align}
   -\frac{\omega^2}{4} +  \frac{3}{4}\widetilde{B}_d(0) = \frac{r_{1c}}{2} - r_c.
\end{align}
Substituting the above condition in Eq.~\eqref{M2}, we obtain
\begin{align}
0 =  r - \frac{r_1}{2} + \frac{r_{1c}}{2} -r_c + \frac{3}{4}u M_0^2.
\end{align}
On solving for $M_0$, we arrive at Eq.~\eqref{Msol} in the main text,
\begin{align}
  |M_0| \propto \sqrt{r_c-r - \biggl(\frac{r_{1c}-r_{1}}{2}\biggr)} \Rightarrow \beta = 1/2.
\end{align}

\section{Retarded Green's Function: Transient behavior} \label{app-grtr}
We now study the one-loop correction to the retarded Green's function in Eq.~\eqref{dGr} taking
into account the transient part of the tadpole Eq.~\eqref{Tdu}.
Using the relation between the driven and the undriven Green's functions Eq.~\eqref{gdu}, the one-loop contribution to $\delta G_R$ is given by
\begin{align}
  \delta G_{R}(k,t,t')&=q^2\cos(\omega t/2)\cos(\omega t'/2)\int_{0}^\infty d\tau \cos^2(\omega \tau/2)\nonumber\\
&\times  G_{0R,u}(\bar{k},t,\tau)\left[iT'(\tau)\right]G_{0R,u}(\bar{k},\tau,t').\label{dGra}
\end{align}
We will discuss the behavior in the limit $\bar{k} t, \bar{k} t'\ll 1$, but with both times long as compared to the cutoff $\bar{\Lambda}t ,\bar{\Lambda} t'\gg 1$. In order to access these limits, we set $k=0$, and use the expression Eq.~\eqref{tadlt} for the tadpole. We also use the identity, $\cos^4(\omega \tau/2) = 3/8 + \cos(\omega \tau)/2 +\cos(2\omega \tau)/8$. This leads to
\begin{align}
&  \delta G_{R}(k=0,t,t')=\frac{8 q}{3}\cos(\omega t/2)\cos(\omega t'/2)\nonumber\\
&\times  \int_{t'}^t d\tau
  \biggl[\frac{3}{8} + \frac{1}{2}\cos(\omega \tau) +\frac{1}{8}\cos(2\omega \tau)\biggr](t-\tau)(\tau-t')
  \frac{\theta}{\tau^2}
  .\label{dGra1}
\end{align}
Now we show that logarithmic corrections begin to emerge when $t \gg t'$. In this limit, the dominant term corresponds to approximating $\tau-t' \rightarrow \tau$. Following this, of the two remaining terms, the
one that dominates at long times is
\begin{align}
&  \delta G_{R}(k=0,t\gg t')= \frac{8 \theta}{3}q\cos(\omega t/2)\cos(\omega t'/2)
\nonumber\\
& \times t\int_{t'}^t d\tau
  \frac{1}{\tau}
  \biggl[\frac{3}{8} + \frac{1}{2}\cos(\omega \tau) +\frac{1}{8}\cos(2\omega \tau)\biggr]\nonumber\\
  &= -G_{0R}(k=0,t \gg t') \frac{8\theta}{3}
  \biggl\{
  \frac{3}{8} \ln\left(t/t'\right)
  \nonumber\\
  &+ \frac{1}{2}\biggl[{\rm Ci}(\omega t)-{\rm Ci}(\omega t')\biggr]
  +\frac{1}{8}\biggl[{\rm Ci}(2\omega t)-{\rm Ci}(2\omega t')\biggr]
  \biggr\},
  \label{dGra2}
\end{align}
where ${\rm Ci}(x) = -\int_{x}^{\infty}dt \frac{\cos{t}}{t}$ is the cosine integral, and we
have used that $G_{0R}(k=0, t\gg t') = - q t \cos(\omega t/2)\cos(\omega t'/2)$. Note that
${\rm Ci}(x) = \gamma + \ln{x}+ O(x)$ for $x\ll 1$ where $\gamma$ is Euler's constant, an observation which will be used below.
At long-times $\omega t \gg 1$, two different limits arise depending on whether
$\omega t' \ll 1$ or $\omega t'\gg 1$. In fact for
$\omega t' \gg 1$, the terms in Eq.~\eqref{dGra1} containing the cosine integral ${\rm Ci}(x)$ do not contribute to any logarithmic corrections, and we find
\begin{equation}
\begin{split}
&\delta G_{R}(k=0,t\gg t', \omega t\gg \omega t' \gg 1)= -G_{0R} \\[2mm]
&\qquad\qquad\times \left[\theta
  \ln\left(t/t'\right) + {\rm finite}\right], \label{dGrb}
\end{split}
\end{equation}
implying that the leading logarithmic singularity of the correction modifies the full Green's function as
\begin{align}
 & G_{R}(k=0,t \gg t', \omega t\gg  \omega t'\gg 1 ) = \nonumber\\
& -q t \cos(\omega t/2)\cos(\omega t'/2)\biggl[1-\theta\ln\bigl(t/t'\bigr)\biggr]\nonumber \\\approx
  &-q t \cos(\omega t/2)\cos(\omega t'/2)\bigl(t'/t\bigr)^{ \theta}.
\end{align}
The above corresponds to Eq.~\eqref{agr1} in the main text.

On the other hand for $\omega t' \ll 1$, but with still $\omega t\gg 1$, the cosine-integral terms in Eq.~\eqref{dGra2} do provide logarithmic contributions and therefore give
\begin{align}
 &   \delta G_{R}(k=0,t\gg t', \omega t \gg 1\gg \omega t')= -G_{0R}(k=0,t\gg t')
    \nonumber\\
    &\times \frac{8\theta}{3}
  \biggl[\frac{3}{8} \ln\left(t/t'\right) -\frac{1}{2}\ln(\omega t') - \frac{1}{8}\ln(2\omega t')\biggr], \label{dGrc}
\end{align}
implying that the leading logarithmic singularity of the correction modifies the full Green's function as
\begin{align}
 & G_{R}(k=0,t \gg t', \omega t\gg 1 \gg \omega t') = \nonumber\\
& \qquad -q t \cos(\omega t/2)\cos(\omega t'/2)\nonumber\\
& \qquad\qquad\times \biggl[1-\theta\ln{(\omega t)}  +  \frac{8\theta}{3} \ln{(\omega t')}\biggr]\nonumber \\\approx
  &-q t \cos(\omega t/2)\cos(\omega t'/2) (\omega t')^{8 \theta/3}/(\omega t)^{\theta}.
\end{align}
The above corresponds to Eq.~\eqref{agr2} in the main text.

\section{Keldysh Green's Function: Transient behavior} \label{app-gktr}

Now we discuss the one loop correction to the Keldysh Green's function Eq.~\eqref{dGk} in the transient regime. Using Eq.~\eqref{gdu} and Eq.~\eqref{Tdu}, we may write the one loop correction Eq.~\eqref{dGk} as
\begin{align}
&\delta iG_{K}(k,t,t')=q^3 \cos(\omega t/2)\cos(\omega t'/2) \nonumber\\
&\times\int_{0}^\infty d\tau \cos^2(\omega \tau/2) G_{0R,u}(\bar{k},t,\tau)\left[iT'(\tau)\right]\ iG_{0K,u}(\bar{k},\tau,t')\nonumber\\ &+\left(t\leftrightarrow t'\right)
\\
&=-\frac{8\theta}{3} q^2 \frac{\sqrt{r_0}}{\bar{k}^3} \cos(\omega t/2)\cos(\omega t'/2)\nonumber\\
&\times \int_0^t d\tau \cos^4(\omega \tau/2) \sin(\bar{k}(t-\tau))(2\bar{\Lambda})^2
\frac{(2\bar{\Lambda} \tau)^2-1}{\left[(2\bar{\Lambda} \tau)^2+1\right]^2}\nonumber\\
&\times \sin(\bar{k}t')\sin(\bar{k}\tau) \nonumber \\&+\left(t\leftrightarrow t'\right),
\end{align}
where $\left(t\leftrightarrow t'\right)$ indicates the same expression but with $t$ and $t'$ exchanged.

We will now use the relation
\begin{align}
 (2\bar{\Lambda})^2\frac{(2\bar{\Lambda} \tau)^2-1}{\left[(2\bar{\Lambda} \tau)^2+1\right]^2}  = -\frac{d}{d\tau}\biggl[\frac{4\bar{\Lambda}^2 \tau}{1 + (2\bar{\Lambda} \tau)^2}\biggr], \label{id1}
\end{align}
in order to integrate by parts, further below, in the integral above.
Thus,
\begin{align}
&  \delta iG_{K}(k,t,t')=\frac{8\theta}{3}  q^2 \frac{\sqrt{r_0}}{\bar{k}^3}\cos(\omega t/2)\cos(\omega t'/2) \nonumber\\
&\times   \int_{0}^t d\tau \cos^4(\omega \tau/2)
  \sin(\bar{k}(t-\tau)) \sin(\bar{k}t')\sin(\bar{k}\tau)\nonumber\\
&\times  \frac{d}{d\tau}\biggl[\frac{4\bar{\Lambda}^2 \tau}{1 + (2\bar{\Lambda} \tau)^2}\biggr] \nonumber\\
  &+ \left(t\leftrightarrow t'\right).
\end{align}

As we did for the one loop correction to the retarded Green's function, we first consider $k=0$ as we are interested in
the behavior at $\bar{k} t,\bar{k} t'\ll 1$. Further, we will explore the same two conditions considered for the retarded Green's functions,
i.e, either $\omega t ,\omega t'\gg 1$, or  $\omega t \gg 1\gg \omega t'$.
We use the identity
$\cos^4(\omega \tau/2) = 3/8 + \cos(\omega \tau)/2 +\cos(2\omega \tau)/8$ to write,
\begin{align}
  &\delta iG_{K}(k=0,t,t')=  \frac{8\theta}{3} q^2 \sqrt{r_0} \cos(\omega t/2)\cos(\omega t'/2)\nonumber\\
&\times  \int_0^t d\tau \biggl[\frac{3}{8} + \frac{1}{2}\cos(\omega \tau) +\frac{1}{8}\cos(2\omega \tau)\biggr](t-\tau)t'\tau\nonumber\\ &\times \frac{d}{d\tau}\biggl[\frac{4\bar{\Lambda}^2 \tau}{1 + (2\bar{\Lambda} \tau)^2}\biggr]\nonumber\\ & +\left(t\leftrightarrow t'\right).
\end{align}
Integrating by parts, we obtain
\begin{align}
&  \delta iG_{K}(k=0,t,t')=  -\frac{8\theta}{3} q^2 \sqrt{r_0} \cos(\omega t/2)\cos(\omega t'/2)\nonumber\\
&\times   \int_0^t d\tau \biggl[\frac{3}{8} + \frac{1}{2}\cos(\omega \tau) +\frac{1}{8}\cos(2\omega \tau)\biggr](t t' - 2 t'\tau)\nonumber\\
&\times \frac{4\bar{\Lambda}^2 \tau}{1 + (2\bar{\Lambda} \tau)^2}
+ \left(t\leftrightarrow t'\right)\nonumber\\
 & +\frac{8\theta}{3} q^2 \sqrt{r_0} \cos(\omega t/2)\cos(\omega t'/2)\nonumber\\
&\times  \int_0^t d\tau \biggl[\frac{\omega}{2}\sin(\omega \tau) +\frac{\omega}{4}\sin(2\omega \tau)\biggr](t t' \tau - t'\tau^2)\frac{4\bar{\Lambda}^2 \tau}{1 + (2\bar{\Lambda} \tau)^2}\nonumber\\
  &+ \left(t\leftrightarrow t'\right).
\end{align}
At times which are large as compared to $\bar{\Lambda}^{-1}$, the dominant terms are those which give logarithmic corrections. These in particular arise from
the first and second lines above. Among them, it is the second term in the numerator $\propto 2 t'\tau, 2 t \tau$ that dominates. Keeping only these dominant terms,
and approximating  $\frac{4\bar{\Lambda}^2 \tau}{1 + (2\bar{\Lambda} \tau)^2}\approx 1/\tau^2$ we obtain
\begin{align}
&  \delta iG_{K}(k=0,t,t')\approx  -\frac{8\theta}{3} q^2 \sqrt{r_0} t t' \cos(\omega t/2)\cos(\omega t'/2)\nonumber\\
&\times  \int_{\bar{\Lambda}^{-1}}^t d\tau \biggl[\frac{3}{8} + \frac{1}{2}\cos(\omega \tau) +\frac{1}{8}\cos(2\omega \tau)\biggr]\frac{1}{\tau} + \left(t\leftrightarrow t'\right).
\end{align}
Performing the integrals, one obtains
\begin{align}
&  \delta iG_{K}(k=0,t,t')\approx  -\frac{8\theta}{3} q^2 \sqrt{r_0} t t'\cos(\omega t/2)\cos(\omega t'/2)\nonumber\\
&\times \biggl\{\frac{3}{8}\ln\bigl(t t' \bar{\Lambda}^2\bigr) + \frac{1}{2}\biggl[{\rm Ci}(\omega t)-{\rm Ci}(\omega/\bar{\Lambda})\biggr] \nonumber\\
  &+\frac{1}{8}\biggl[{\rm Ci}(2\omega t)-{\rm Ci}(2\omega/\bar{\Lambda})\biggr]
  + \frac{1}{2}\biggl[{\rm Ci}(\omega t')-{\rm Ci}(\omega/\bar{\Lambda})\biggr]
  \nonumber\\
&  +\frac{1}{8}\biggl[{\rm Ci}(2\omega t')-{\rm Ci}(2\omega/\bar{\Lambda})\biggr]\biggr\}.
\end{align}

For $\omega t\gg 1, \omega t' \gg 1$ the terms involving the cosine-integral ${\rm Ci}(x)$  do not give any logarithmic corrections. Noting that the Green's function at the Gaussian level is
\begin{align}
i G_{0K}(k=0, t, t') = q^2 \sqrt{r_0}\cos(\omega t/2)\cos(\omega t'/2) \,\, t t',
\end{align}
the leading logarithmic singularities and the correction discussed above modify the interacting Green's function at one-loop as
\begin{align}
 &\!\! i G_K(k=0, \omega t \gg 1 ,\omega t' \gg 1)\!\! = q^2 \sqrt{r_0} \cos(\omega t/2)\cos(\omega t'/2)
\nonumber\\
&\times t t'\biggl[1 -
    \theta \ln\bigl(\bar{\Lambda}^2 t t'\bigr)\biggr]\\
  &\propto q^2 \sqrt{r_0}  \cos(\omega t/2)\cos(\omega t'/2)  \bigl(\bar{\Lambda}^2 t t'\bigr)^{1 - \theta}.
\end{align}
The above is Eq.~\eqref{gktr1} in the main text.

On the other hand if $\omega t \gg 1 \gg \omega t'$, then, some of the cosine-integrals do contribute with a logarithm singularity, giving
    \begin{align}
&  \delta iG_{K}(k=0,\omega t \gg 1 \gg \omega t')\approx  -\frac{8\theta}{3} q^2 \sqrt{r_0} t t'\nonumber\\
&\times \cos(\omega t/2)
\cos(\omega t'/2)\biggl[\frac{3}{8}\ln(\bar{\Lambda} t)+ \ln(\bar{\Lambda} t')\biggr],
\end{align}
where we have used that ${\rm Ci}(\omega/\bar{\Lambda}) \approx \ln(\omega/\bar{\Lambda})$ for $\omega \ll \bar{\Lambda}$.
Thus
\begin{align}
&  i G_K(k=0, \omega t \gg 1 \gg \omega t') = q^2 \sqrt{r_0} \cos(\omega t/2)\cos(\omega t'/2)\nonumber\\
&\times t t'\biggl[1 -
    \theta \ln(\bar{\Lambda}t) - \frac{8 \theta}{3} \ln(\bar{\Lambda}t')\biggr]\\
  &\approx q^2 \sqrt{r_0}  \cos(\omega t/2)\cos(\omega t'/2)  (\bar{\Lambda }t)^{1- \theta}
  (\bar{\Lambda}t')^{1-8\theta/3}.
\end{align}
The above is Eq.~\eqref{gktr2} in the main text.

\section{Keldysh Green's function: Long-time behavior} \label{app-gklt}

At the Gaussian level, the equal-time Keldysh Green's function is
\begin{align}
  iG_{0K}(k, t,t) = q^2 \cos^2(\omega t/2) \frac{\sqrt{r_0}}{\bar{k}^2} \sin^2(\bar{k}t).
\end{align}
We will now study the one-loop correction to this quantity in the limit of $\bar{k} t \gg 1$.
Using Eq.~\eqref{dGk} and Eq.~\eqref{Tdu} along with the identity Eq.~\eqref{id1}, we obtain
\begin{align}
&  \delta iG_{K}(k,t,t)=\frac{16\theta}{3}  q^2 \frac{\sqrt{r_0}}{\bar{k}^3}\cos^2(\omega t/2) \nonumber\\
&\times   \int_{0}^t d\tau \cos^4(\omega \tau/2)
  \sin(\bar{k}(t-\tau)) \sin(\bar{k}t)\sin(\bar{k}\tau)\nonumber\\ &\times \frac{d}{d\tau}\biggl[\frac{4\bar{\Lambda}^2 \tau}{1 + (2\bar{\Lambda} \tau)^2}\biggr].
\end{align}
Above when we expand $\sin(\bar{k}(t-\tau))$ we only keep the $\sin(\bar{k} t) \cos(\bar{k}\tau)$ term
as this is the term that is proportional to the Gaussian Green's function $G_{0K}$.
Thus
\begin{align}
&  \delta i G_K (k, t,t) \approx \frac{8 \theta}{3} q^2 \frac{\sqrt{r_0}}{\bar{k}^3}\sin^2(\bar{k}t)\cos^2(\omega t/2)\nonumber\\
&\times \int_{0}^t d\tau \cos^4(\omega \tau/2)
  \sin(2\bar{k}\tau)\frac{d}{d\tau}\biggl[\frac{4\bar{\Lambda}^2 \tau}{1 + (2\bar{\Lambda}\tau)^2}\biggr]\nonumber\\
  &=\frac{8\theta}{3}  q^2 \frac{\sqrt{r_0}}{\bar{k}^3}\sin^2(\bar{k}t)\cos^2(\omega t/2)\nonumber\\
&\times   \biggl[\cos^4(\omega t/2) \sin(2\bar{k}t)\frac{4\bar{\Lambda}^2 t}{1 + (2\bar{\Lambda} t)^2}
  \nonumber\\
&-2\bar{k}\int_{0}^t d\tau \cos^4(\omega \tau/2)\cos(2\bar{k}\tau) \frac{4\bar{\Lambda}^2 \tau}{1 + (2\bar{\Lambda} \tau)^2}\nonumber\\
  &
  + 2\omega \int_{0}^t d\tau \cos^3(\omega \tau/2)\sin(\omega \tau/2)\cos(2\bar{k}\tau) \frac{4\bar{\Lambda}^2 \tau}{1 + (2\bar{\Lambda} \tau)^2}\biggr].
\end{align}
For $\bar{\Lambda} t \gg 1$, i.e., for times that are long as compared to the microscopic scale $\bar{\Lambda}^{-1}$, the first term falls off at long times as $1/t$, and therefore we drop it.
Collecting the remaining terms, we obtain
\begin{align}
&    \delta i G_K (k, t,t) \approx \frac{8\theta}{3}  q^2 \frac{\sqrt{r_0}}{\bar{k}^3}\sin^2(\bar{k}t)\cos^2(\omega t/2)\nonumber\\
&\times \biggl\{-2\bar{k}\int_{{\bar{\Lambda}^{-1}}}^t d\tau \biggl[\frac{3}{8} + \frac{1}{2}\cos(\omega \tau) +\frac{1}{8}\cos(2\omega \tau)\biggr]
    \cos(2\bar{k}\tau) \frac{1}{\tau}\nonumber\\
    &+ \omega \int_{{\bar{\Lambda}^{-1}}}^t d\tau
    \biggl[\frac{1}{2} + \frac{1}{2}\cos(\omega \tau)\biggr]\sin(\omega \tau)
    \cos(2\bar{k}\tau) \frac{1}{\tau}
    \biggr\}.
\end{align}
Note that the second term does not give any logarithmic contributions.
We then define $T(\alpha)$ for $\alpha \ll \bar{\Lambda}$
\begin{align}
 T(\alpha)&= \int_{{\bar{\Lambda}^{-1}}}^{t} d\tau \frac{\cos(\alpha \tau)}{\tau} = {\rm Ci}(\alpha t)- {\rm Ci}(\alpha/\bar{\Lambda})\nonumber\\
& \approx -\gamma -\ln(\alpha/\bar{\Lambda}) + {\rm Ci}(\alpha t),
\end{align}
to write
\begin{align}
 & \delta i G_K (k, t,t)  \approx -\frac{16\theta}{3}  q^2 \frac{\sqrt{r_0}}{\bar{k}^2}\sin^2(\bar{k}t)\cos^2(\omega t/2)\nonumber\\
&\times  \biggl[\frac{3}{8}T(2\bar{k}) + \frac{1}{4} T(2\bar{k}+\omega)
  +\frac{1}{4} T(2\bar{k}-\omega)\nonumber\\ &+\frac{1}{16}T(2\bar{k}+2\omega)+\frac{1}{16} T(2\bar{k}-2\omega)\biggr].
\end{align}
Now at times longer than $\alpha^{-1}$, where $\alpha=2\bar{k}, \bigl|2\bar{k} \pm \omega\bigr|$, and $\bigl|2\bar{k} \pm 2\omega\bigr|$, the ${\rm Ci}(\alpha t)$ terms can be dropped  because they decrease as a power-law in $\alpha t$. Thus at long-times we obtain
\begin{align}
& \delta i G_K (k, t,t) \approx iG_{0K}(k,t,t) \frac{16  \theta}{3}\biggl[ \frac{3}{8}\ln(\bar{k}/\bar{\Lambda})
\nonumber\\
&+ \frac{1}{4} \ln((2\bar{k}+\omega)/\bar{\Lambda})
  +\frac{1}{4} \ln((2\bar{k}-\omega)/\bar{\Lambda})\nonumber\\ &+\frac{1}{16}\ln((2\bar{k}+2\omega)/\bar{\Lambda})
  +\frac{1}{16} \ln((2\bar{k}-2\omega)/\bar{\Lambda}) \biggr].
\end{align}
In the long wavelength limit
$\bar{k} \ll \omega$, only the first logarithm dominates in the correction above. Therefore the resulting $G_K$ can be written as
\begin{align}
iG_K(k, t, t; \bar{k} \ll \omega) \approx q^2\frac{\sqrt{r_0}}{\bar{k}^{2-2  \theta}}\sin^2(\bar{k} t)\cos^2(\omega t/2).
\end{align}
The above is Eq.~\eqref{gktr3} in the main text.

\section{Derivation of Eq.~\eqref{GR1x}}\label{app-ft}

In order to determine the asymptotic behavior of the Green's functions discussed in Sec.~\ref{sec:spatial}, it is sufficient to study the terms with $m=\pm 1$ in Eq.~\eqref{IRK}, as the behavior for $m=\pm 2$ follows in a straightforward manner.
In addition, Eqs.~\eqref{IRK} have terms that are resonant due to their denominators being proportional to $\bar{k}-m\omega/2$. In doing the Fourier transform in space for highlighting the structure of the spatial correlations, these resonant terms dominate over the non-resonant terms. Accordingly, the analysis below focuses only on the resonant contributions.

Using Eq.~\eqref{eq:ft}, the spatial Fourier transform  in $d=4$ requires computing
\begin{align}
  {\cal I}_{R,K}(x,t,t') = \frac{1}{4 \pi^2 x} \int_0^{\Lambda} dk k^2 J_{1}(k x) I_{R,K}(k,t,t'),
\end{align}
where ${\cal I}_{R,K} (x,t,t')$ contribute to the one loop corrections $G_{R,K}^{(1)}$ and
\begin{align}\label{iprk}
    I_{R}(k,t,t') &= \sum_{m=\pm 1}\cos(m\omega(t+t')/2)\nonumber\\
&\quad\times \frac{\sin((k-m\omega/2)(t-t'))}{k^2(k-m\omega/2)},\\
    I_K(k,t,t) &= \sum_{m=\pm 1}\frac{1}{k^3(2k-m\omega)}\nonumber\\
&\quad\times    \biggl\{[1+\cos(m \omega t)][1-\cos(2kt)]
 \nonumber \\
&\quad-\sin(2kt)\sin(m\omega t)\biggr\}.
\end{align}
Above we have kept only the resonant terms in Eq.~\eqref{IRK}. For notational convenience, we have also set $v=1$, so that $\bar{k}=k$.

\subsection{Evaluation of ${\cal I}_R$}
In order to evaluate ${\cal I}_R$, we perform a shift of variables $k -m\omega /2 \rightarrow k$ and obtain
\begin{align}
  {\cal I}_{R}(x,t,t') &=
  \cos(\omega(t+t')/2) \frac{1}{4 \pi^2 x} \nonumber\\
  &\times \biggl[\int_{-\omega/2}^{\Lambda-\omega/2}dk \frac{\sin(k(t-t'))}{k} J_1(k x + \omega x/2)\nonumber\\
  &+ \int_{\omega/2}^{\Lambda+\omega/2}dk \frac{\sin(k(t-t'))}{k} J_1(k x - \omega x/2)\biggr]
  \nonumber\\
  &\approx \cos(\omega(t+t')/2) \frac{1}{4 \pi^2 x}\nonumber\\
  &\times \biggl\{\int_{-\omega/2}^0 dk \frac{\sin(k(t-t'))}{k} J_1(k x + \omega x/2)\nonumber\\
    &+ \int_{\omega/2}^0 dk \frac{\sin(k(t-t'))}{k} J_1(k x - \omega x/2) \nonumber\\
    & +\int_0^{\Lambda} dk \frac{\sin(k(t-t'))}{k}\biggl[J_1(k x + \omega x/2) \nonumber\\
&    + J_1(k x - \omega x/2) \biggr]
    \biggr\},
    \label{eqIPR-1}
\end{align}
where in the last term we have used that $\Lambda \gg \omega$.
In the first term of Eq.~\eqref{eqIPR-1}, we perform the
transformation $k \rightarrow -k$ and use that $J_1(-x) = - J_1(x)$ to write
\begin{align}
&  {\cal I}_{R}(x,t,t') =  \cos(\omega(t+t')/2) \frac{1}{4 \pi^2 x}\nonumber\\
&\times  \biggl\{- 2\int^{\omega/2}_0 dk \frac{\sin(k(t-t'))}{k} J_1(k x - \omega x/2)
    \nonumber\\
    & +\int_0^{\Lambda} dk \frac{\sin(k(t-t'))}{k}\biggl[J_1(k x + \omega x/2) \nonumber\\
&    + J_1(k x - \omega x/2) \biggr]
    \biggr\}.
\end{align}
Following this, we use that $J_1$ may be replaced by its asymptotic value $J_1(k x + \omega x/2) =-\sqrt{\frac{2}{\pi (k x + \omega x/2)}}\cos(k x + \omega x/2+\pi/4)$ for $\omega x\gg 1$.
This gives
\begin{align}
&  {\cal I}_{R}(x, t,t') \approx \cos(\omega(t+t')/2)\sqrt{\frac{2}{\pi}} \frac{1}{4 \pi^2 x^{3/2}}
  \nonumber\\
&\times  \biggl\{2\int^{\omega/2}_0 dk \frac{\sin(k(t-t'))}{k} \frac{\cos(kx - \omega x/2 + \pi/4)}{\sqrt{k-\omega/2}}
    \nonumber\\
    & -\int_0^{\Lambda} dk \frac{\sin(k(t-t'))}{k}\biggl[\frac{\cos(k x +\omega x/2 + \pi/4)}{\sqrt{k+\omega/2}}\nonumber\\
&    +  \frac{\cos(k x - \omega x/2 + \pi/4)}{\sqrt{k-\omega/2}}\biggr]
    \biggr\}.\label{IPR}
\end{align}
We now discuss the two non-trivial cases, i.e., the behavior on the light-cone and the behavior inside it.

On the light-cone, since $x=t-t'$, (and $t>t'$ due to the causal structure of the retarded Green's function)
we write
\begin{align}
&  {\cal I}_{R}(x= t-t') \approx \cos(\omega(t+t')/2)\sqrt{\frac{2}{\pi}} \frac{1}{4 \pi^2 x^{3/2}}
\nonumber\\
&\times  \biggl\{2\int^{\omega/2}_0 dk \frac{\sin(k x)}{k} \frac{\cos(kx - \omega x/2 + \pi/4)}{\sqrt{k-\omega/2}}
    \nonumber\\
    & -\int_0^{\Lambda} dk \frac{\sin(k x)}{k}\biggl[\frac{\cos(k x +\omega x/2 + \pi/4)}{\sqrt{k+\omega/2}}\nonumber\\
 &   +  \frac{\cos(k x - \omega x/2 + \pi/4)}{\sqrt{k-\omega/2}}\biggr]
    \biggr\} \nonumber\\
  &\approx \cos(\omega(t+t')/2)\sqrt{\frac{2}{\pi}} \frac{1}{4 \pi^2 x^{3/2}}\nonumber\\
&\times   \biggl\{ 2\sin(\omega x/2 - \pi/4)\int^{\omega/2}_0 dk \frac{\sin^2(k x)}{k} \frac{1}{\sqrt{k-\omega/2}}
    \nonumber\\
    & -\int_0^{\Lambda} dk \frac{\sin^2(k x)}{k}\biggl[-\frac{\sin(\omega x/2 + \pi/4)}{\sqrt{k+\omega/2}}\nonumber\\
&    +  \frac{\sin(\omega x/2 - \pi/4)}{\sqrt{k-\omega/2}}\biggr]
    \biggr\}.
\end{align}
In the last equality of the previous equation we have only kept terms that oscillate in phase.
Since this expression is dominated by $k$ away from $\pm \omega/2$ we may write
\begin{align}
&  {\cal I}_{R}(x= t-t') \approx \cos(\omega(t+t')/2)\sqrt{\frac{2}{\pi}} \frac{x^{1/2}}{4 \pi^2 x^{3/2}}\nonumber\\
&\times  \biggl\{2\sin(\omega x/2 - \pi/4)\int^{\omega x/2}_0 dy \frac{\sin^2(y)}{y} \frac{1}{\sqrt{y}}
    \nonumber\\
    & -\int_0^{\Lambda x} dy  \frac{\sin^2(y)}{y}\biggl[-\frac{\sin(\omega x/2 + \pi/4)}{\sqrt{y}}\nonumber\\
&    +  \frac{\sin(\omega x/2 - \pi/4)}{\sqrt{y}}\biggr]
    \biggr\}.
\end{align}
Performing the $y$-integral we find
  \begin{align}
    {\cal I}_{R}(x= t-t') &\propto \cos(\omega(t+t')/2) \sin(\omega x/2 + \delta)\frac{1}{x^{3/2}},
    \label{irlc1}
  \end{align}
where $\delta$ stands for a constant phase shift that originates from the $\pi/4$ phase in the asymptotic expansion of the Bessel function.

For studying the behavior of ${\cal I}_R$ at points away from the light cone, let us introduce $y = k(t-t')$. Then Eq.~\eqref{IPR} becomes
\begin{align}
&{\cal I}_{R}(x,t,t') =  \cos(\omega(t+t')/2) \frac{1}{4 \pi^2 x}\nonumber\\
&\times \biggl\{- 2\int^{\omega (t-t')/2}_0 dy \frac{\sin(y)}{y} J_1(y x/(t-t') - \omega x/2)
  \nonumber\\
  &+\int_0^{\Lambda (t-t')} dy \frac{\sin(y)}{y}\biggl[J_1(y x/(t-t') + \omega x/2) \nonumber\\
&  + J_1(y x/(t-t') - \omega x/2) \biggr]
    \biggr\}.
\end{align}
When $x \ll t-t'$, and since $\omega (t-t'), \Lambda(t-t') \gg 1$, the upper limits of integration in the expression above can be set to $\infty$.  We can also approximate
$J_1(y x/(t-t') + \omega x/2) \approx J_1(\omega x/2)$.
Moreover, in the
second term above, the two Bessel functions cancel each other giving
\begin{align}
  {\cal I}_{R}(x\ll t-t') &=  \cos(\omega(t+t')/2) \frac{1}{4 \pi^2 x}\nonumber\\
&\times J_{1}(\omega x/2)  \biggl(2\int^{\infty}_0 dy \frac{\sin y}{y}\biggr).
\end{align}
Accordingly, inside the light-cone, we obtain
  \begin{align}
    {\cal I}_{R}(x\ll t-t') &= \cos(\omega(t+t')/2) \frac{1}{4 \pi x}J_{1}(\omega x/2) \nonumber\\
  &  \propto \cos(\omega(t+t')/2) \sin(\omega x/2 + \widetilde{\delta})\frac{1}{x^{3/2}},\label{irlc2}
  \end{align}
where in the last line we have used the asymptotic form of $J_1(\omega x/2)$ and $\widetilde{\delta}$ denotes a constant phase, which as before, originates from the $\pi/4$ phase coming from the asymptotic expansion of the Bessel function.
Thus, together with Eq.~\eqref{irlc1} and Eq.~\eqref{irlc2}, we have recovered the asymptotic form
reported in Eq.~\eqref{GR1x} for the retarded Green's function.

\subsection{Evaluation of ${\cal I}_K$}

We now derive the asymptotic behavior of the equal-time Keldysh Green's function at one loop.
Using Eq.~\eqref{iprk} and assuming $\Lambda \gg \omega$, we obtain
\begin{align}
 {\cal I}_{K}(x,t,t) &= \frac{1}{4\pi^2 x}\int_{-\omega/2}^{\Lambda}dk
 \frac{1}{2 k} \biggl\{\sin(2 k t)\sin(\omega t) \nonumber\\
& + [1-\cos(2 k t)][1+\cos(\omega t)]\biggr\}\frac{J_1(k x + \omega x/2)}{k+\omega/2}\nonumber\\
    &+ \frac{1}{4\pi^2 x}\int_{\omega/2}^{\Lambda}dk
  \frac{1}{2 k} \biggl\{-\sin(2 k t)\sin(\omega t) \nonumber\\
&  + [1-\cos(2 k t)] [1+\cos(\omega t)]\biggr\}\frac{J_1(k x - \omega x/2)}{k-\omega/2}.
\end{align}
As before, we split the integral as follows:
\begin{align}
&  {\cal I}_{K}(x,t,t) = \frac{1}{4\pi^2 x}\int_{-\omega/2}^0 dk \frac{1}{2 k}
  \biggl\{\sin(2 k t)\sin(\omega t) \nonumber\\
  &+ [1-\cos(2 k t)] [1+\cos(\omega t)]\biggr\}\frac{J_1(k x + \omega x/2)}{k+\omega/2}\nonumber\\
  &+\frac{1}{4\pi^2 x}\int_0^{\omega/2}dk\frac{1}{2k}\biggl\{\sin(2 k t)\sin(\omega t)
  \nonumber\\
  &- [1-\cos(2 k t)][1+\cos(\omega t)]
    \biggr\}\frac{J_{1}(k x -\omega x/2)}{k-\omega/2}\nonumber\\
  & + \frac{1}{4\pi^2 x}\int_0^{\Lambda}dk\frac{1}{2k}\biggl\{\frac{J_1(k x +\omega x/2)}{k+\omega/2}
    \biggl[\sin(2 k t)\sin(\omega t)\nonumber\\
    &+ (1-\cos(2 k t)) (1+\cos(\omega t))\biggr]
    \nonumber\\
 &+ \frac{J_1(k x - \omega x/2)}{k-\omega/2}\biggl[ -\sin(2 k t)\sin(\omega t) \nonumber\\
& + (1-\cos(2 k t)) (1+\cos(\omega t))\biggr]  \biggr\}.
\end{align}
Performing the transformation $k \rightarrow -k$ in the first term and noting that $J_1(-x) = -J_1(x)$, we obtain
\begin{align}
&  {\cal I}_{K}(x,t,t) =\frac{1}{2\pi^2 x}\int_0^{\omega/2}dk\frac{1}{2k}\biggl\{\sin(2 k t)\sin(\omega t) \nonumber\\ &- [1-\cos(2 k t)] [1+\cos(\omega t)]
    \biggr\}\frac{J_{1}(k x -\omega x/2)}{k-\omega/2} \nonumber\\
  &+\frac{1}{4\pi^2 x}\int_0^{\Lambda}dk\frac{1}{2k}\biggl\{\frac{J_1(k x +\omega x/2)}{k+\omega/2}
    \biggl[\sin(2 k t)\sin(\omega t) \nonumber\\&+ (1-\cos(2 k t)) (1+\cos(\omega t))\biggr]
    \nonumber\\
 &+ \frac{J_1(k x - \omega x/2)}{k-\omega/2}\biggl[ -\sin(2 k t)\sin(\omega t)\nonumber\\& + (1-\cos(2 k t)) (1+\cos(\omega t))\biggr] \biggr\}.\label{IPK}
\end{align}
We now separately discuss the behavior on the light-cone and inside the light-cone.
On the light-cone, since $x=2t$, we write Eq.~\eqref{IPK} as
\begin{align}
&  {\cal I}_{K}(x=2t) =\frac{1}{2\pi^2}\int_0^{\omega t}dy\frac{1}{2y}\biggl\{\sin y \sin(\omega t)
\nonumber\\
&- (1-\cos y)[1+\cos(\omega t)]
    \biggr\}\frac{J_{1}(y -\omega x/2)}{y-\omega x/2} \nonumber\\
  &+\frac{1}{4\pi^2}\int_0^{\Lambda x}dy\frac{1}{2y}\biggl\{\frac{J_1(y +\omega x/2)}{y+\omega x/2}
\biggl[\sin y \sin(\omega t) \nonumber\\
&    + (1-\cos y ) (1+\cos(\omega t))\biggr]\nonumber\\
 &+ \frac{J_1(y - \omega x/2)}{y-\omega x/2}\biggl[ -\sin y \sin(\omega t) \nonumber\\
& + (1-\cos y) (1+\cos(\omega t))\biggr]   \biggr\}.
\end{align}
Replacing $J_1$ by its leading asymptotic form  we obtain
\begin{align}
& {\cal I}_{K}(x=2t) =-\frac{1}{2\pi^2}\sqrt{\frac{2}{\pi}}\int_0^{\omega t}dy\frac{1}{2y}\biggl\{\sin y \sin(\omega t) \nonumber\\
&- (1-\cos y) [1+\cos(\omega t)]
    \biggr\}\frac{\cos(y -\omega x/2+\pi/4)}{(y-\omega x/2)^{3/2}} \nonumber\\
  &-\frac{1}{4\pi^2}\sqrt{\frac{2}{\pi}}\int_0^{\Lambda x}dy\frac{1}{2y}\biggl\{\frac{\cos(y +\omega x/2+\pi/4)}{(y+\omega x/2)^{3/2}}
    \biggl[ \sin y \sin(\omega t) \nonumber \\
    &+ (1-\cos y ) (1+\cos(\omega t))\biggr]
    \nonumber\\
 &+ \frac{\cos(y - \omega x/2+\pi/4)}{(y-\omega x/2)^{3/2}}\biggl[ -\sin y \sin(\omega t)\nonumber\\
& + (1-\cos y ) (1+\cos(\omega t))\biggr]   \biggr\}.
\end{align}
This equation implies
  \begin{align}
    {\cal I}_{K}(x=2t) &\propto \frac{\cos(\omega x/2 +\alpha'')}{x^{3/2}},\label{ipklc1}
   \end{align}
where $\alpha''$ is a constant phase-shift originating from the $\pi/4$ phase in the asymptotic expansion of the Bessel function.

Now we discuss the behavior inside the light-cone, corresponding to having $2t \gg x$. Here we may replace $1-\cos(2 k t) \approx 1$, obtaining
\begin{align}
  {\cal I}_{K}(2t\gg x) &\approx  \frac{1}{4\pi^2}\int_1^{\Lambda x}dy\frac{1}{2y}\biggl[\frac{J_1(y +\omega x/2)}{y+\omega x/2}\nonumber\\
&    +\frac{J_1(y - \omega x/2)}{y-\omega x/2} \biggr]
  \nonumber\\
  & -\frac{1}{2\pi^2}\int_1^{\omega x/2} dy \frac{1}{2y} \frac{J_1(y - \omega x/2)}{y-\omega x/2}.
\end{align}
Let us assume $\omega x \gg 1$, so that we can set both the upper-limits of integration $\Lambda x =\omega x=\infty$. Then
\begin{align}
  {\cal I}_{K}(2t\gg x) &\approx  \frac{1}{4\pi^2}\int_1^{\infty}dy\frac{1}{2y}\biggl[\frac{J_1(y +\omega x/2)}{y+\omega x/2}\nonumber\\
&    -\frac{J_1(y - \omega x/2)}{y-\omega x/2} \biggr]\nonumber\\
  &  \propto \frac{\cos(\omega x/2 + \alpha''')}{x^{3/2}}.\label{ipklc2}
\end{align}
Above $\alpha'''$ is a constant phase-shift, which as before can be traced back to the $\pi/4$ phase in the asymptotic form of the Bessel function. Equations~\eqref{ipklc1} and \eqref{ipklc2} together summarize the asymptotic behavior of the Keldysh Green's function at one loop, and yield Eq.~\eqref{GR1x} anticipated in the main text.

%

\end{document}